\documentstyle[aps,preprint]{revtex}
\input epsf

\tightenlines

\global\firstfigfalse  

\def\rem#1{}
\def\bra#1{\langle #1 |}
\def\ket#1{| #1\rangle}

\def\OO{{\cal O}}
\def\KK{\Xi}
\def\KK{{\cal P }}
\def\NN{{\cal N}}
\def\TT{{\cal T}}
\def\QQ{{\cal Q}}

\def\none{$\NN=1$}
\def\nonestar{$\NN=1^*$}

\def\nfour{$\NN=4$}

\def\half{{1\over2}}

\def\bel{\begin{equation}\label}
\def\ee{\end{equation}}
\newcommand\eref[1]{(\ref{#1})}
\newcommand\Eref[1]{Eq.~\eref{#1}}
\newcommand\vev[1]{\left\langle{#1}\right\rangle}

\newcommand{\OL}[1]{ \hspace{1pt}\overline{\hspace{-1pt}#1
   \hspace{-1pt}}\hspace{1pt} }

\begin{document}

\baselineskip=20pt

\pagestyle{empty}
\preprint{
\begin{minipage}[t]{3in}
\begin{flushright} NSF-ITP-02-139,
UW/PT 02-23
\\
hep-th/0209211
\\[5pt]
\hphantom{.}
\end{flushright}
\end{minipage}
}

\title{Deep Inelastic Scattering and Gauge/String Duality}
\author{Joseph Polchinski\thanks{Institute for Theoretical Physics,
University of California, Santa Barbara CA 93106-4030}  \ and
Matthew J. Strassler\thanks{Dept. of Physics,
University of Washington, Seattle, WA 98195}}
\maketitle
\begin{abstract} We study deep inelastic scattering in gauge theories which have dual string 
descriptions.  As a function of $gN$ we find a transition.  For small
$gN$, the dominant operators in the OPE are the usual ones, of
approximate twist two, corresponding to scattering from weakly
interacting partons.  For large $gN$, double-trace operators dominate,
corresponding to scattering from entire hadrons (either the original
`valence' hadron or part of a hadron cloud.)  At large $gN$ we
calculate the structure functions.  As a function of Bjorken $x$ there
are three regimes: $x$ of order one, where the scattering produces
only supergravity states; $x$ small, where excited strings are
produced; and, $x$ exponentially small, where the excited strings are
comparable in size to the AdS space.  The last regime requires in
principle a full string calculation in curved spacetime, but the
effect of string growth can be simply obtained from the world-sheet
renormalization group.

\end{abstract}

\newpage

\section{Introduction}

The discovery of gauge/string duality~\cite{maldacon} has given new
insight into both gauge theory and string theory.  In this paper we use
the duality to study a gauge theory process, deep inelastic scattering
(DIS), which has played an important role in the history of the strong
interaction.  This process probes the internal structure of hadrons,
and so should distinguish a field theory, where there are pointlike
constituents, from a string theory, where there are not. It is
therefore interesting to see how this is reconciled in gauge theories
that have a weakly coupled string description, and how the physics
evolves as we interpolate from such a theory to one that has a weakly
coupled gauge theory description at high energy.

Our work is directed at a better understanding both of gauge theory and
of string theory.  First, it gives a new perspective on the
field-theoretic analysis of DIS.  Second, we hope that it will shed
some light on the possible form of a string dual to QCD, extending our
earlier work on elastic scattering~\cite{PShard}.  Third, we find that
on the string side a complete analysis requires us to develop some new
methods for calculating string amplitudes in curved spacetime, which
may be useful in other contexts.

We will focus on confining gauge theories that are scale invariant, or
nearly so, at momenta well above the confinement scale $\Lambda$.  A
key distinction is whether the high energy scale-invariant theory is
weakly or strongly coupled.  Standard weakly coupled examples include
asymptotically free theories such as $SU(N)$ Yang-Mills, QCD, and
\none\ supersymmetric Yang-Mills.  In these, scale-invariance is
violated by logarithms, but in any given momentum range above
$\Lambda$ the coupling $\alpha(\mu)\equiv g_{YM}^2/4\pi$ is nearly
constant.  In the strongly coupled case, gauge theory perturbation
theory is not useful at any scale, but in there may be a weakly
coupled string dual.  It is particularly interesting to study examples
with a parameter that allows us to move continuously between the two
regimes.

A model that one might bear in mind is \nonestar\
supersymmetric Yang-Mills theory~\cite{rdew,PS} (though our presentation
will be more general).  This is \nfour\ supersymmetric Yang-Mills, a
conformal theory, explicitly broken at a scale $m$ to pure \none\
Yang-Mills.  The massless fields are those of an asymptotically-free
confining gauge theory, but at the scale $m$ there are massive scalars
and fermions in the adjoint representation, which do not affect
confinement but do regulate the ultraviolet of the theory.  In
particular, $\alpha(\mu)$ is constant for $\mu>m$ and runs below $m$.
If $\alpha N$ is large in the ultraviolet then the theory is conformal
down to the scale $m\sim \Lambda$; if $\alpha N$ is small in the
ultraviolet then the theory is conformal down to the scale $m$, and
then runs logarithmically down to the scale $\Lambda\sim m
e^{-2\pi/3\alpha(m) N}$.  Pure \none\ Yang-Mills is restored for
$m\to\infty$, $\alpha(m)\to 0$, $\Lambda$ fixed.

\begin{figure}[h]
\begin{center}
\leavevmode
\epsfbox{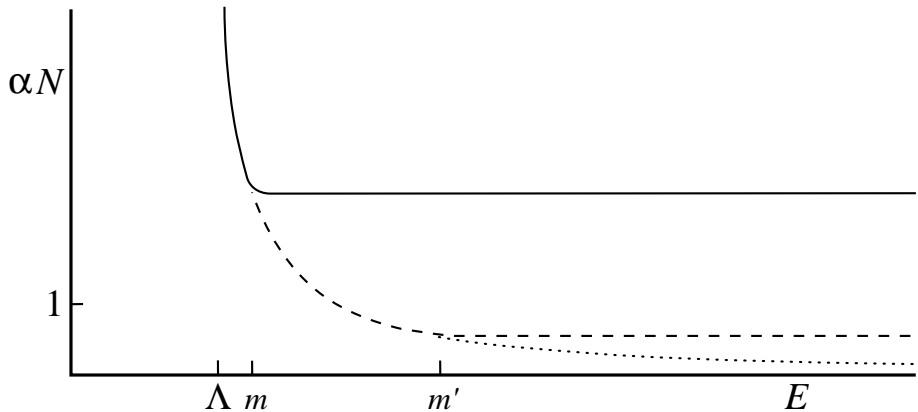}
\end{center}
\caption{Running coupling in the \nonestar\ theory.  For the solid
curve the asymptotic $\alpha N$ is large and the decoupling scale $m$ is
close to $\Lambda$.  For the dashed curve the asymptotic $\alpha N$ is
small and the decoupling scale is $m' \gg\Lambda$.  The asymptotically
free \none\ curve is dotted.}
\end{figure}
When $m\gg\Lambda$, the dual string coupling $g=\alpha$ is small but
the space on which the strings propagate is highly curved,
with curvature radius $R\sim (gN)^{1/4}$, so there is no
weakly-coupled string description.  Fortunately, in this case ordinary
field theoretic perturbation theory works well for $\mu>\Lambda$, both
below and above $\mu=m$.   Conversely, in order that $g N$ be
large (with $g$ small), so as to allow a weakly-coupled string
description, the scale $m$ of the additional matter must be
essentially the confinement scale $\Lambda$ where the mass gap is
generated.  In this case field theory perturbative techniques are
not useful at any scale.  By varying $g$ we can interpolate between
these two cases.

The analysis in Sections II and III is field theoretic.
In Section II we review DIS, including the definitions of the kinematic
variables $q^2$ and $x$.  We also review the operator product expansion
(OPE) analysis, and apply it at small 't Hooft coupling.  Section III
extends the OPE analysis to large 't Hooft coupling.

The results are as
follows.  At small 't Hooft coupling, where field theory perturbation
theory applies, hadronic substructure is similar to that in QCD. Each
hadron typically has a small number of partons carrying most of the
energy, surrounded by a cloud of wee partons with very small
momentum fraction $x$.  At large 't Hooft coupling, where the stringy
dual description is perturbative, we find that hadronic substructure is
qualitatively different.  All the partons are wee; parton evolution is
so rapid that the probability of finding a parton with a substantial
fraction of the energy is vanishingly small.  At finite $N$ each hadron
has a diffuse cloud of other hadrons surrounding it.  For most
processes the incoming electron is most likely, as  $q^2\to \infty$ 
with fixed $N$
and
$x$, to scatter off one of the hadrons in the
cloud.  It is as though the electron could find no quarks at finite $x$
inside the proton, due to their fragmentation into huge numbers of
quarks, antiquarks and gluons, and instead were more likely to strike
the very diffuse pion cloud around the proton.

The transition from one behavior to the other is most
likely continuous,
and takes the following form: the traditional lowest-twist
operators of QCD develop large anomalous dimensions and become
high-twist operators as the 't Hooft coupling becomes large.  However,
certain double-trace operators, normally subleading in QCD, do not get
large anomalous dimensions; their twists remain relatively low.
Consequently, these traditionally higher-twist operators begin to
dominate deep-inelastic scattering as the 't Hooft coupling grows
large.  From the string theory side, the transition is most easily
understood beginning from large 't Hooft coupling, when the background
on which the string propagates is weakly curved. As the 't Hooft
coupling shrinks and the curvature radius of the background becomes
small, the masses of stringy modes shrink relative to those of the
Kaluza-Klein modes.  When both modes have masses of the same order, a
cross-over to new behavior occurs.  This
transition is the same one which connects the picture of
D-branes, with open strings coupled to closed strings, to that of
black holes, with only closed strings.

Sections IV and V present the string theoretic calculations of the DIS
structure functions.  In terms of calculational method we find that
there are three distinct regimes of $x$: of order one, of order
$(gN)^{-1/2}$, and of order $e^{-(gN)^{1/2}}$.  Section IV deals with
the first of these.  In this regime excited strings are not produced,
and so the calculation involves only supergravity degrees of freedom.
The results are consistent with the earlier OPE analysis, and in
addition give the detailed form of the structure functions.  We
discuss the qualitative form of the amplitudes, where the the warped
spacetime geometry plays an important role, and we compare the
inelastic results here to the results for the elastic amplitudes in
Ref.~\cite{PShard}.

To complete the analysis --- in particular, to verify the
energy-momentum sum rule --- we must determine the amplitudes also at
small $x$.  This is done in Section \mbox{V}.  In this case the
scattering produces excited strings.  We first analyze the amplitude
assuming, as in Ref.~\cite{PShard}, that the scattering process is
localized and so can be approximated locally by the flat spacetime
amplitude.  The resulting structure functions give Regge behavior (consistent
with a Pomeron-exchange picture) but with a divergent momentum
sum rule.  The effect that cuts this off is the logarithmic growth of
strings at high energy~\cite{growth}.  When $x$ is exponentially
small, the string size becomes comparable to the AdS radius and
existing methods for perturbative string calculation break down.  By
using the world-sheet renormalization group we are able to include the
effects of string growth.  This sums all orders in $(gN)^{-1/2} \ln x$
and produces a result valid down to $x = 0$.

Section VI presents conclusions and speculations.

We have learned that O.~Andreev is also considering DIS in 
gauge/gravity duality.
We also note the recent papers~\cite{newhard} on elastic scattering in the
hard and Regge regimes.

\section{Deep Inelastic Scattering }

Deep inelastic scattering --- the scattering of electrons
off of hadrons in kinematic regimes where the hadron is
broken apart --- is a natural probe of hadronic substructure.
The electron plays no role except to emit an off-shell
photon of four-momentum $q^\mu$; the photon then strikes
the hadron, probing it near the lightcone at distances
of order $1/\sqrt {q^2}$.
As shown by Bjorken, if hadrons are made of
essentially free massless partons which appear in the hadronic
wave function with some distribution of momentum
and energy, then this distribution can be measured.
In particular, defining $x= -q^2/2P\cdot q$, where
$P^\mu$ is the momentum of the scattered hadron, the
probability of finding a parton with four-momentum $xP^\mu$
is a distribution function $f(x,q^2)$ that is independent of $q^2$.

This $q^2$-independence, called {\it Bjorken scaling}, is a property
only of the truly free parton model.  In real QCD, scaling is of
course violated; the parton distribution functions ``evolve'' as
$q^2$ increases, because each parton, through QCD interactions,
tends to split into multiple partons of smaller $x$.
Consequently the apparent structure of a QCD hadron depends on $q^2$,
with the number of partons increasing and their average $x$
decreasing as $q^2$ increases.  This physical picture can be derived from
a precise operator product analysis, which we now review.

\rem{Generally it contains some quarks and antiquarks, and
a few gluons, at moderate $x$, but has many gluons, along with some
quarks and antiquarks, at small $x$, with the number of
gluons increasing as $q^2$ grows.  The large number of
partons at very small $x$ were called {\it wee partons} by
Feynman.}
\rem{This is in fact due to two effects, generally
intermingled in standard textbook treatments on the subject, but
properly separated in early work by Kogut and Susskind \cite{kosu}.  In
particular, violations of Bjorken scaling occur (1) through anomalous
dimensions of certain operators (which could be present even in a
scale-invariant theory) and (2) through running of the gauge coupling.
In perturbation theory, these both lead to logarithmic violations of
Bjorken scaling.  However, this is somewhat misleading.  The
logarithms from the running of the gauge coupling are real logarithms
of momentum scale $q^2$; the gauge coupling runs logarithmically.  But
the logarithms from the anomalous dimensions are in fact {\it powers}
of $q^2$.}

\subsection{Review of general formalism}

We will use the conventions and notations of Ref.~\cite{manohar}, 
except that we
take the metric $(-1,+1,+1,+1)$ to connect with standard string 
calculations.  As
explained in any standard treatment, the DIS amplitudes for electron-hadron
scattering can be extracted from the imaginary part of forward Compton
scattering, or more precisely from the matrix element of two
electromagnetic currents\footnote{The following formalism can also be
applied with minor modifications in the cases of nonabelian symmetry
currents and of the energy-momentum tensor.  The latter is
particularly useful, since some of the theories to which we would like
to apply this formalism --- for example, pure Yang-Mills theory --- do
not have spin-one currents. In such a case, our results concerning
hadronic structure will still apply, although they have to be
extracted formally from graviton-hadron scattering.}  inside the hadron
of interest
\bel{Wdef}
{\sf T}^{\mu\nu} \equiv i
\int\ d^4y\, e^{iq\cdot y} \bra{P,\QQ} T(J^\mu(y)J^\nu(0)) \ket{P,\QQ}\ .
\ee
Here $T(\OO_1,\OO_2)$ indicates a time-ordered product, $P^\mu$ is the
momentum of the hadron, and $\QQ$ is its electromagnetic charge.
In complete generality this can be written
\bel{Wexpand}
{\sf T}_{\mu\nu}= \widetilde F_1\left(x,{q^2\over \Lambda^2}\right)
\left(g_{\mu\nu}
-{q_\mu q_\nu\over q^2}\right) +
{2x\over q^2}\widetilde  F_2\left(x,{q^2\over\Lambda^2}\right)
\left(P_\mu + {q_\mu \over 2x}\right)
\left(P_\nu + {q_\nu \over 2x}\right) \ .
\ee
Here
\bel{bjx}
x=-q^2/2P\cdot q\ ,\quad q^2 > 0\ ,\quad 0 \leq x \leq 1\ .
\ee
$\Lambda$ is a mass-scale in QCD ---
typically the hadron mass
$M^2 = -P^2$.  Note that the structure functions $\widetilde F_1 $ and
$\widetilde F_2$ are dimensionless; in the parton model their imaginary parts
are related to the parton distribution functions, as will be discussed below.
These functions may be extracted from $g^{\mu\nu}{\sf T}_{\mu\nu}$ and
$P^\mu P^\nu {\sf T}_{\mu\nu}$; note that $q^\mu  {\sf T}_{\mu\nu}$ vanishes
by current conservation.

In the unphysical region of $x\gg 1$, ${\sf T}^{\mu\nu}$ can be reexpressed
using the operator product expansion (OPE) of the two currents.
On general grounds [conservation of $J^\mu$ and Lorentz
invariance, and dropping terms that would vanish in the diagonal matrix
element~(\ref{Wdef}),] the OPE takes the form 
\begin{eqnarray}\label{genOPE}
\int d^4y\ e^{iq\cdot y}\,  T(J^\mu(y)J^\nu(0)) &=&
\sum_{n=0,2,4,\dots}\sum_j
    \OO_{n,j}^{\rho_1\ldots \rho_n}(0)
    {q_{\rho_3} \ldots q_{\rho_n}\over
q^{n + \delta_{n,j}}} \Biggl( \frac{\Lambda^2}{q^2}
\Biggr)^{\gamma_{n,j}/2}
\nonumber \\
& & \qquad \Bigg[
\left(q^2 g_{\mu\nu} -
{q_\mu q_\nu} \right) {q_{\rho_1} q_{\rho_2}}  C_{n,j}^{(1)} \nonumber \\
& & \qquad\quad +
  \left( q^2 g_{\mu\rho_1} - {q_\mu
q_{\rho_1}} \right)
\left(q^2 g_{\nu\rho_2} - {q_\nu q_{\rho_2}}\right)
    C_{n,j}^{(2)} \  \Bigg] \ .
\end{eqnarray}
Here $n$ is the spin of the operator $\OO_{n,j}$ and $j$ indexes the
various operators of spin $n$; the engineering dimension of
$\OO_{n,j}$ is $\delta_{n,j}$ and $\gamma_{n,j}$ is its anomalous
dimension.  We also define $\Delta_{n,j} = \delta_{n,j} +
\gamma_{n,j}$ as the total scaling dimension of $\OO_{n,j}$ and
$\tau_{n,j}=\Delta_{n,j}-n$ is its twist.  For simplicity we have
written this for the scale-invariant case, where the dimensions
$\Delta_{n,j}$ are constant but in general noncanonical.  In QCD the
dimensions vary only logarithmically, through the running of the
coupling, and the above expression can be used locally in $q^2$.  The
$C_{n,j}^{(s)}$ are dimensionless.

By Lorentz invariance, the (spin-averaged) matrix
elements take the form
\bel{matel}
\bra{P,\QQ}\OO_{n,j}^{\rho_1\cdots\rho_{n}}\ket{P,\QQ}
=
\Lambda^{\delta_{n,j} - n - 2} (-2)^n P^{\rho_1}\cdots P^{\rho_{n}}
{A}_{n,j} + {\rm traces}\ ,
\ee
where ${A}_{n,j}$ is a pure
number.
The factors of $-2P^\rho$ combine
with those of $q_\rho$ to give factors of
$-2P\cdot q = q^2/x$, and so
\begin{eqnarray}
{\sf T}^{\mu\nu} &\approx& i
\sum_{n\ {\rm even}}\sum_j x^{-n}
\Biggl({\Lambda^2\over q^2}\Biggr)^{\half\tau_{n,j}-1}\
\nonumber \\
& &\qquad \left\{\Biggl( g_{\mu\nu} -
{q_\mu q_\nu\over q^2}\Biggr){A}_{n,j} C^{(1)}_{n,j} +{4x^2\over q^2}
\biggl(P_\mu + {q_\mu \over 2x}\biggr)
\biggl(P_\nu + {q_\nu \over 2x}\biggr) {A}_{n,j} C^{(2)}_{n,j}
    \right\}\ ,
\end{eqnarray}
where we have dropped the trace terms, which are suppressed by powers of
$q^2$. Thus
\bel{Fs}
\widetilde F_s(x,q^2) \approx i
\sum_{n\ {\rm even}} \sum_j C^{(s)}_{n,j} A_{n,j} \, x^{-n}(2x)^{s-1}
\left({\Lambda^2\over q^2}\right)^{\half\tau_{n,j}-1}\ \ \ (s=1,2)\ .
\ee
We see that the operators that dominate the amplitudes at large $q^2$
are those that have the smallest {\it twist} $\tau$.
Note both ${C}_{n,j}^{(s)}$ and
${A}_{n,j}$ depend on the normalization of $\OO_{n,j}$; however only
the product ${C}_{n,j}^{(s)}{A}_{n,j}$ appears in the end.

The OPE determines the behavior in the limit that $q^2$ and $x$ are
both large.  The physical region for DIS is $0 \leq x \leq 1$.  A
standard contour argument relates the behavior at large $x$ to the
moments of the DIS structure functions.  Integrate with respect to
$\omega\equiv 1/x$ around a contour surrounding $\omega=0$
\bel{intFs}
{1\over 2\pi}\int_{|\omega|=\epsilon} d\omega\ \omega^{-n-1}
(\omega/2)^{s-1}
\widetilde F_s(\omega^{-1},q^2)
\approx { C^{(s)}_{n,j} A_{n,j}}
\Biggl({\Lambda^2\over q^2}\Biggr)^{\half\tau_{n,j}-1}\ .
\ee
One now deforms the contour out to $|\omega|=\infty$, except along
the real axis for $\omega>1$ and $\omega<-1$, where there are branch
cuts.  The optical theorem for the discontinuity across the branch cut
then determines the moments
\bel{Ms}
M^{(s)}_n(q^2)
\equiv  \int_0^1 dx\ x^{n-1} (2x)^{1-s}{F_s(x,q^2)}
\approx \frac{1}{4} \sum_j{ C^{(s)}_{n,j} A_{n,j}}
\left({\Lambda^2\over q^2}\right)^{\half\tau_{n,j}-1}\
\ (n\ {\rm even}, s=1,2)
\ee
where
\bel{ftf}
F_s(x,q^2) \equiv 2\pi\, {\rm Im}\, \widetilde F_s(x,q^2)\ .
\ee
These structure functions are the standard ones appearing in
the hadronic tensor
\bel{Wdefb}
{\sf W}^{\mu\nu} \equiv i
\int\ d^4y\, e^{iq\cdot y} \bra{P,\QQ} [J^\mu(y),J^\nu(0)] \ket{P,\QQ}\ .
\ee
The second term in the commutator vanishes, and the first, with a
complete set of states inserted between the currents, gives the square
of the DIS amplitude.  The functions $F_s$ are defined by
\bel{Wexpandb}
{\sf W}_{\mu\nu}=  F_1\left(x,{q^2\over \Lambda^2}\right)
\left(g_{\mu\nu}
-{q_\mu q_\nu\over q^2}\right) +
{2x\over q^2} F_2\left(x,{q^2\over\Lambda^2}\right)
\left(P_\mu + {q_\mu \over 2x}\right)
\left(P_\nu + {q_\nu \over 2x}\right) \ .
\ee

\subsection{Specializing to weak coupling}

In a free parton model (QCD or Yang-Mills at zero coupling) it is easy
to see that the only operators that appear in the $JJ$ OPE have
$\tau=2, 4, 6, \ldots$. In a theory with adjoint fields only, each
operator involves a certain number of traces over color indices.  The
leading multi-trace operators have $\tau\geq 4$, so DIS is dominated
by a set of single-trace twist-2 operators $\TT_{n,j}$, with classical
dimension $n+2$ and spin $n$.  These include the energy-momentum
tensor.\footnote{If the theory has quark fields in the fundamental
representation, then there are quark-antiquark bilinear operators; the
notion of multi- and single-trace operators generalizes to operators
which can or cannot be factored into gauge-invariant suboperators.}

For finite coupling, the $\TT_{n,j}$ develop
anomalous dimensions $\gamma_{n,j}$ (which are positive)  except
for the energy-momentum tensor whose conservation always implies
$\gamma=0$.
Of course, in leading-order perturbation theory the anomalous
dimensions $\gamma_{n,j}\sim \alpha N$ ($N$ is the number of colors and
$\alpha$ is the coupling at the scale $q^2$) so in an asymptotically-free
theory such as QCD or Yang-Mills, the $\TT_{n,j}$ have twist close to
2 at high $q^2$ and are the lowest-twist operators appearing in the
OPE.

For theories weakly coupled in the UV, the moments are then
\bel{MsAF}
M^{(s)}_n(q^2)
\approx \frac{1}{4} {\sum_j}' { C^{(s)}_{n,j} A_{n,j}}
\left({\Lambda^2\over q^2}\right)^{\half\tau_{n,j}-1}\
\approx \frac{1}{4} {\sum_j}' { C^{(s)}_{n,j} A_{n,j}}
\left({\Lambda^2\over q^2}\right)^{\half\gamma_{n,j}}\
\ \ (n\ {\rm even}, s=1,2)\ ,
\ee
where the prime on the sum indicates that we keep only the terms
corresponding to the $\TT_{n,j}$; other terms are
suppressed by powers of $q^2$.
The $\gamma_{n,j}$ are positive, so the $M_n$ decrease to zero as $q^2$
increases. The exception is $n=2$: the energy-momentum tensor has
no anomalous dimension and therefore gives
$q^2$-independent sum rules
\begin{eqnarray}\label{mom1}
M_2^{(1)}(q^2) &\equiv& \int_0^1 dx\, x
    \, F_1(x,q^2)\ {\to}\ {\rm constant} \ ,
\\[4pt]
  \label{mom2}  M_2^{(2)}(q^2)&\equiv& \half \int_0^1 dx\,
F_2(x,q^2)\ \to\ {\rm constant}\ ,
\end{eqnarray}
as $q^2 \to \infty$.
In general, however,
\bel{mom3}
M_n^{(s)}\propto
\left({\Lambda^2\over q^2}\right)^{\half\tau_{n,j}-1} ,\quad
\tau_n\equiv\min_j{\tau_{n,j}}\ee
at large $q^2$.
The fact that the moments vary as powers of $q^2$ in a conformal
field theory was noted in ref.~\cite{kosu}; we will therefore call
this power-law violation of Bjorken scaling
{\it Kogut-Susskind evolution.}

\subsection{Parton interpretation of leading-twist effects}

In a parton model, one interprets ${\sf W}^{\mu\nu}$ in terms of
distributions of partons inside the hadron.  If the partons have
spin-$\half$ one finds that $F_2 = 2xF_1$ (the Callan-Gross relation) and
that
\bel{fdef}
F_1(x,q^2) = \half \sum_i {\cal Q}_i^2 f_i(x)\ .
\ee
Here $f_i$ is the parton distribution function of parton-type $i$,
which has charge ${\cal Q}_i$. Thus both $F_1$ and $F_2$ are
$q^2$-independent in the parton model; this is Bjorken scaling.

Since in real QCD the anomalous dimensions of the $\TT_{n,j}$ are not
zero, the functions $F_1$ and $F_2$ evolve with $q^2$.  This is
expressed through the DGLAP equations \cite{DGLAP}, which may be
written
\bel{DGLAP}
q^2 {dM_n\over dq^2} = -\half\gamma_n M_n
\ee
in the simple case where there is only one $\TT_{n,j}$ for each $n$.
This is understood conceptually as evolution of the parton
distribution functions with $q^2$, through the relation between the
$F_s$ and the $f_i$.  In leading-order perturbation theory the
$\gamma_n$ are simply the moments of the parton splitting functions
(times a factor of $\alpha_s$).  The positivity of the $\gamma_n$
ensures that evolution acts to decrease the energy-fraction $x$ of the
average parton, as one would expect on general physical grounds: as
$q^2$ increases, the functions $f_i(x,q^2)$ tend to shrink at larger
$x$ and grow at small $x$.  This is of course observed experimentally
in QCD.\footnote{QCD formulas are often written in a way which
obscures the connection with conformal field theory. That
Kogut-Susskind evolution occurs locally in $q^2$ in QCD can be seen
by expanding standard QCD expressions for \eref{mom3} and \eref{DGLAP}
in the one-loop beta function coefficient $b_0$ around a fiducial
momentum scale $q_0^2$; this allows a separation of running coupling
effects, $b_0\ln (q^2/q_0^2)$, from the power laws
$q^{-\gamma (\alpha(q_0^2))}$.}

The
energy-momentum sum
rule from $M_2$,
\bel{mom4}
M_2=\frac{1}{2}\int_0^1 dx\ \sum_i x \QQ_i^2 f_i(x,q^2) \to {\rm constant}
\ee
simply states in the parton model that the total charge-weighted energy of all
partons  does not change as one probes the system at increasingly short
distances.

\section{Strong coupling: field theory analysis}

\subsection{The OPE at large 't Hooft coupling}

Let us now apply this formalism to deep-inelastic scattering
in four-dimensional
confining theories in which the
't Hooft coupling  $g N$ can be varied between large and small values,  and
which are in either case nearly conformal above the confinement scale as in
Fig.~1.
We will assume that the theory has a symmetry
current
whose associated charge is carried by light degrees of freedom which
are present inside hadrons (we will consider $U(1)$ currents, but again
nonabelian currents or the energy momentum tensor could also be 
used).   We can then study the
current-current matrix element, \Eref{Wdef}, in the hadron of
interest.
Note that the hadrons might only be
dipoles under this $U(1)$; for example, baryon number in QCD is not
carried by any meson but a photon coupling to baryon number will still
scatter off partons inside the meson.

We saw above that a generic theory at
small $g N$ has DIS physics similar
to that of QCD. At large $q^2$, hadrons can be treated
as bound states of weakly-interacting partons with accompanying
distribution functions.  Since the anomalous dimensions of all
single-trace operators are of order $g N$, the $\tau\approx 2$
operators $\TT_{n,j}$ dominate the physics.  The moments $M_n$ of the
functions $F_1$ and $F_2/2x$ vary slowly, and Bjorken scaling is only
weakly violated.

However, at large $g N$, the physics is totally different.  One learns
from AdS/CFT duality~\cite{maldacon,GKPW} that the operators
$\TT_{n,j}$ (excepting the energy-momentum tensor) have {\it large}
anomalous dimensions.  These operators are related to states in the IIB string
spectrum, and their dimensions (and consequently their anomalous
dimensions and their twists) are of order\footnote
{Ref.~\cite{GKP2} has recently discussed these operators in AdS/CFT duality,
noting that for large spin ($n>\sqrt{gN}$)
their anomalous dimensions are small compared to
their spins.  However, the anomalous dimensions are still large compared to
one, so they are not the leading contribution to DIS.}   
\bel{bigdim}
\Delta \sim \tau\sim \gamma \sim ( gN)^{1/4}\gg 1 \ .
\ee
So large are these anomalous dimensions that (excepting $T^{\mu\nu}$)
the $\TT_{n,j}$ are no longer the leading-twist operators. On general
grounds, there are {\it double-trace} operators which do not receive
large anomalous dimensions for any $g N$.  It is these operators which
dominate the OPE and are lowest-twist at large $g N$.

What are these double-trace operators?  First, note that any theory,
supersymmetric or not, has {\it single-trace} operators whose
anomalous dimensions are order 1 even at large $g N$.\footnote{Here
we consider single-
and double-trace operators constructed out of adjoint fields only, as
in the original $AdS_5 \times S^5$ duality~\cite{maldacon} and its simplest
variants.  For large-$N$ QCD, the currents may couple to fields in the
fundamental representation, and in place of a trace we would have a
quark-antiquark bilinear (possibly with adjoint fields in between).  The
ensuing analysis is parallel, replacing $N$ by $N^{1/2}$ in the discussion
that begins with \Eref{Cs}.}  Let us call these operators {\it protected}
and refer to them as $\KK_p$ (with dimension $\Delta_p$, spin $n_p$, twist
$\tau_p=\Delta_p-n_p,$ and charge $\QQ_p$ under the $U(1)$ symmetry)
where the index $p$ simply labels the operators. The energy-momentum
tensor is such an operator; so are conserved currents; in
supersymmetric theories other examples would include, but are not
restricted to \cite{gubser}, the chiral operators.

  Now we may construct double-trace operators, as bilinears in the
$\KK_p$, which can appear in the $JJ$ OPE.  For simplicity, let us
consider a spin-zero operator $\KK_p$, and consider its bilinears, of
the form $\KK_p^{\dagger}\KK_p^{\vphantom\dagger}$, $\KK_p^{\dagger}
\partial_\mu\partial_\nu\KK_p^{\vphantom\dagger}$, and so
on. Large-$N$ factorization as $N\to\infty$ implies that
\bel{KKtwopt}
\vev{\KK_p^\dagger \KK_p^{\vphantom\dagger}(y)\,
\KK_p^\dagger \KK_p^{\vphantom\dagger}(0) }=
\left| \vev{\KK_p^{\vphantom\dagger} (y)\,
\KK_p^\dagger(0) }\right|^2
+ O\left({1\over N^2}\right)\
\sim r^{-4\Delta_p\pm{\rm order}\left(1/ N^2\right)  }\ ,
\ee
and thus the twist of $\KK^\dagger \KK$ is (at least) twice that of
$\KK$, up to a correction of order $1/N^2$.  Similar arguments imply
that the operators $\KK^\dagger \partial_{\rho_1}\partial_{\rho_2}
\cdots\partial_{\rho_r}\KK$ have $\Delta = r + 2\Delta_p$ and
$\tau\geq 2\tau_p$, up to an ${\rm order}\left(\pm{1/ N^2}\right)$
correction.  When all Lorentz indices are symmetrized, with traces
removed, to make an operator of maximum spin, the twist takes the
minimum value $\tau \approx 2\tau_p$. Since $\KK_p$ is protected,
$\Delta_p$, and consequently $\tau_p$, has a finite limit as $g
N\to\infty$. Altogether we conclude these double-trace operators have
finite $\tau$ as $gN\to\infty$, and thus have smaller twist than any
$\TT_{n,j}$ at large $gN$.

This is by no means the complete set of double-trace operators which
may appear in the OPE.  We may also build them from
$\KK_p$ of non-zero spin, in which case index contractions are more
complex.  Furthermore, if $\KK_p$ and $\KK_{p'}$ have the same global
charges, then operators such as
$\KK_p^{\dagger}\partial_{\mu}\partial_\nu\KK_{p'}$ may
appear.  Although these operators are not subleading to the ones
mentioned above, and must be included in all applications, we will
omit them in our formulas in order to keep our presentation simple.

    The essential point of this discussion is that these operators
behave differently at large $gN$ from the unprotected operators
$\TT_{n,j}$.  While the single-trace operators $\TT_{n,j}$ have
anomalous dimensions of order $g N$ in perturbation theory, and of
order $(g N)^{1/4} \to \infty$ for $g N\to\infty$, {\it a double-trace
operator constructed from protected single-trace operators is itself
protected: its anomalous dimension is largely inherited from its
constituent single-trace operators}, up to a shift of order $g /N$ in
perturbation theory and of order $N^{-2}h(g N)$ for any $g N$. The
function $h(g N)$ can in principle be computed for small and for large
$g N$.

\rem{It is easy to show that in a free gauge theory with no
gauge-invariant fields, all gauge-invariant operators
have twist 2 or greater.  This means that all double-trace
operators of the form above have $\tau\geq 4$.
There is a unitarity bound that, in a conformal
field theory, all gauge-invariant operators
must have twist 1 or greater.  Those which have
twist 1 must be decoupled from the rest
of the theory.  We will consider more generally theories
for which the operators $\KK_p$ have twist greater
than or equal to a lower bound $\tau_0$ which is
greater than 1 by an amount of order 1, as in the
case just discussed, so that all double-trace
operators have $\tau-2\sim>  1$.  The case where $\tau_0=1$ will
be considered separately below.}

When we include both the $\TT_{n,j}$ and the
$\KK_p^\dagger\partial\partial\cdots\partial\KK_p$ from the OPE,
the moments~(\ref{MsAF}) become
\bel{Msschem}
M^{(s)}_n(q^2)\approx \frac{1}{4} {\sum_j}'
{ C^{(s)}_{n,j} A_{n,j}}
\left({\Lambda^2\over q^2}\right)^{\half\tau_{n,j}-1}\
+\frac{1}{4}{\sum_{p}}'' {C^{(s)}_{n,p}  A_{n,p}}
\left({\Lambda^2\over q^2}\right)^{\tau_p-1}\ ,
\ee
where the primed sum runs over the $\TT_{n,j}$ and the double-primed sum over
the protected double-trace operators with spin $n$ and dimension
$\approx 2\tau_p+n$.

For small $g N$, $\tau_{n,j}\sim 2
  + {\rm order}{( g N)} < 2\tau_p$; for large $g N$,
$\tau_{n,j} \sim (g N)^{1/4}$ but $\tau_p$ remains finite and order
1.  Therefore, the first term dominates as $q^2\to \infty$ for small
$gN$ while the last term dominates for large $gN$.\footnote{There are
special cases which require a slightly different treatment.  For
example, if there is a
protected operator $\KK_{p_0}$ such that $\tau_{p_0} = 1$ at weak
coupling --- e.g., if there is a point particle which
couples to both photons and hadrons ---
then $\KK^\dagger_{p_0}\KK^{\vphantom\dagger}_{p_0}$ is also
a $\TT_{n,j}$ and should only be counted once in~\Eref{Msschem}.
The adjustments to our formulation in this and other special cases
is straightforward; our results on hadronic structure
are not affected.} {\it Thus there is a qualitative
transition at $gN\sim 1$ in which double-trace operators
become the lowest-twist operators in the theory,}  aside from
the energy-momentum
tensor which remains twist-two.

Before interpreting this transition physically, we do the $N$ counting for
the different contributions to the moments.  The leading planar amplitude is
of order $N^2$, and we normalize the currents and other single-trace operators
such that they create hadrons at order $N^0$, and so have two-point functions
of order $N^0$.  (Note that with this normalization the partons have charges of
order
$N^{-1}$; our $U(1)$ currents must be multiplied by a factor of $N$ to give
the usual normalization for the
$R$-currents of \nonestar\ and similar theories.) 
Then for the OPE coefficients we have
\bel{Cs}
   C_{n,j}^{(s)}
\propto \vev{J\ J\ \TT_{n,j}}\sim{N^{-1}}  \ ; \quad
C_{n,p}^{(s)}\propto \vev{J\ J\ \KK_p^\dagger(\partial)^r\KK_p}
\sim {N^{-2}}\ .
\ee
For the matrix elements
\bel{HTH}
   A_{n,j} = \bra{\QQ,P}\TT_{n,j}\ket{\QQ,P}\sim N^{-1}
\ ;
\ee
\bel{HOOHa}
    A_{n,p} =
\bra{\QQ,P} \KK_p^\dagger(\partial)^r\KK^{\vphantom{\dagger}}_p\ket{\QQ,P}
\sim
{N^0}
{\rm \ if }  \ \bra{\QQ,P}\KK^{\vphantom{\dagger}}_p\ket{0}\neq 0\ ;
\ee
\bel{HOOHb}
A_{n,p}=
\bra{\QQ,P} \KK_p^\dagger(\partial)^r\KK^{\vphantom{\dagger}}_p\ket{\QQ,P}
\sim {N^{-2}}
{\rm \ if }  \ \bra{\QQ,P}\KK^{\vphantom{\dagger}}_p\ket{0}=0\ .
\ee
The last two equations follow from the fact that the
matrix element is dominated by a (dis)connected graph if
$\bra{\QQ,P}\KK_p\ket{0}$ is (non)zero.

Thus, from \Eref{HTH}--\eref{HOOHb},
almost all of the matrix elements $A_{n,p}$
appearing in the second sum in \eref{Msschem} are
of order $1/N^2$.  Only for those $\KK_p$ that
have charge $\QQ$ are the elements (potentially)
of order 1.  Let us therefore write $A_{n,p} = a_{n,p}/N^2$ for those
$\KK_p$ with $\QQ_p\neq \QQ$, and separate the
operators by charge:
\begin{eqnarray}\label{Msschemb}
M^{(s)}_n(q^2)&\approx& \frac{1}{4} {\sum_j}'
{ C^{(s)}_{n,j} A_{n,j}}
\left({\Lambda^2\over q^2}\right)^{\half\tau_{n,j}-1}\
+\frac{1}{4} {\sum_{\QQ_p = \QQ}}'' {C^{(s)}_{n,p}  A_{n,p}}
\left({\Lambda^2\over q^2}\right)^{\tau_p-1}
\nonumber \\
&&\qquad +{1\over 4N^2}  {\sum_{\QQ_p \neq \QQ}}'' {C^{(s)}_{n,p}a_{n,p}}
\left({\Lambda^2\over q^2}\right)^{\tau_p-1}\ .
\end{eqnarray}
  From this we see that while the first term dominates at small $g N$,
the situation at large $g N$ is more complex. At $q^2$ somewhat larger
than $\Lambda^2$ the second term always dominates; but it falls as
$(q^2)^{-\tau_\QQ+1}$, where $\tau_\QQ$ is the minimum twist of the
operators $\KK_p$ of charge $\QQ$.  Typically $\tau_\QQ\sim \QQ$; for
example, if all fields carry electric charge 0 or 1, this will be the
case.  If $\tau_\QQ\neq \tau_{\rm c}$ (where $\tau_{\rm c}$ is the
minimum twist of {\it all} electrically charged\footnote{If $\KK_p$ is
neutral then the OPE coefficient $C^{(s)}_{n,p}$ is suppressed.} operators
$\KK_p$) then there is yet another transition.  When $q^2 \agt
\Lambda^2 N^{2/(\tau_\QQ-\tau_{\rm c})}$, the third term becomes the
largest of the three: the contribution of the operator $\KK_p$ with
lowest twist ($\tau_p=\tau_{\rm c}$) falls only as $(q^2)^{-\tau_{\rm
c}+1}$, and thus overcomes its overall $1/N^2$ suppression to dominate
the amplitude as $q^2\to-\infty$.

\begin{figure}
\begin{center}
\leavevmode
\epsfbox{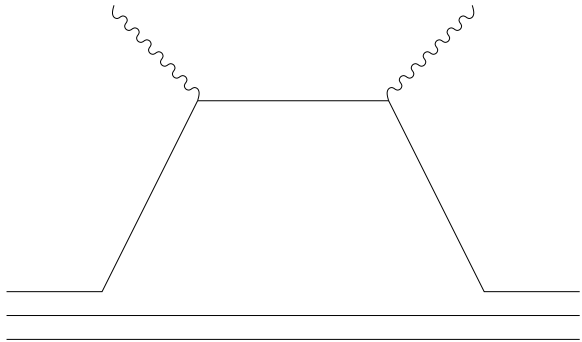}
\end{center}
\caption{Forward Compton amplitude in the parton model.}
\end{figure}

\subsection{Interpretation of the various contributions}

Now we turn to the interpretation of this transition.  The essential
point is that the operators $\TT_{n,j}$ are associated with partons,
while the $\KK_p^\dagger(\partial)^r\KK^{\vphantom \dagger}_p$ are
associated with hadrons.  The local operators $\TT_{n,j}$ couple to
the internal constituents of hadrons (in QCD they are bilinear in
quarks), while at large $N$ the operator
$\KK_p^\dagger(\partial)^r\KK^{\vphantom \dagger}_p$ destroys and
creates a whole hadron, and so is like a bilinear in hadron fields.


Let us first address the large anomalous dimensions of the
$\TT_{n,j}$.  Consider the DGLAP equations \eref{DGLAP}, combined with
the parton model expression \eref{fdef} and the fact that anomalous
dimensions $\gamma_n$ are proportional to the moments of the parton
splitting amplitudes.  From these we see that the $\gamma_{n,j}$
controls the rate of parton splitting.  That $\gamma_{n,j}\gg 1$ at
large $gN$ implies that parton-splitting processes, and consequent
evolution of the parton distribution functions, are vastly more rapid
at large $gN$.  The contributions of $\TT_{n,j}$ to the moments
$M_n^{(s)}$ ($n>2$) decrease rapidly as the partons in the hadron
split repeatedly, leaving parton distribution functions which only
have support at very small $x$ (just how small, we will investigate in
section~V).  This behavior explains why the first term in
\Eref{Msschemb} is so suppressed at large $gN$.

\begin{figure}[h]
\begin{center} 
\leavevmode
\epsfbox{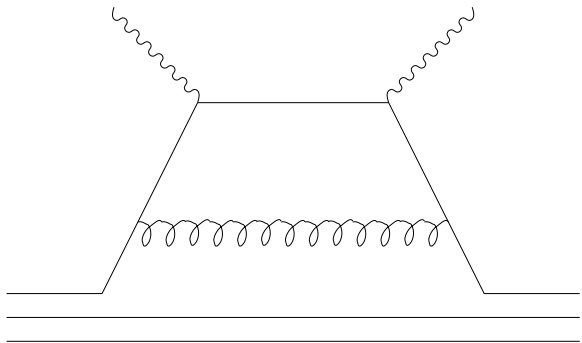}
\end{center}
\caption{Forward Compton amplitude with parton evolution; the
parton splits, losing some fraction of its energy, 
before being struck by the photon.}
\end{figure}

\begin{figure}[h]
\begin{center}
\leavevmode
\epsfbox{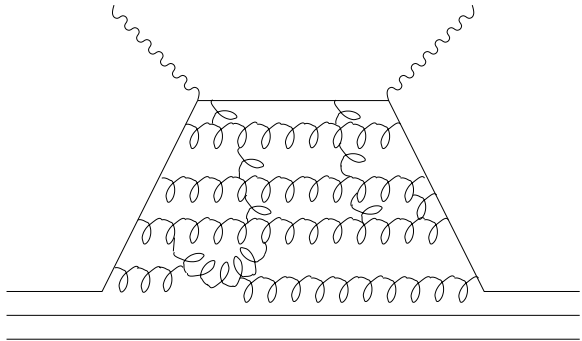}
\end{center}
\caption{As $gN$ increases the splitting rate increases as well; a 
large fraction of the parton's energy is lost before it is struck.}
\end{figure}

Once the partons have fragmented into tiny pieces, what remains to
dominate the moments $M_n^{(s)}$?  If the currents cannot scatter
off of partons, then perhaps they can scatter off of entire hadrons.
Unlike partons, which carry color and radiate strongly
at large $gN$, colorless hadrons are much less likely to lose
their energy through radiation.  In fact, at $N\to\infty$, for any $gN$,
they cannot do so at all.  In this limit, the only thing which can happen
at moderate values of $x$ is that the currents can scatter off
the entire parent hadron $\ket{\QQ,P}$.  One might naively expect
that this contribution would have support only at $x=1$.  However,
as we will calculate shortly, this is not so, because the scattering
need not be elastic even though individual partons are not struck.
Instead the currents can coherently excite the internal structure of
the hadron, without breaking it.

\begin{figure}[h]
\begin{center}
\leavevmode
\epsfbox{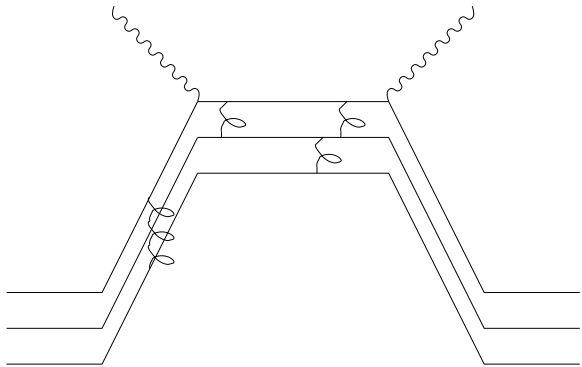}
\end{center}
\caption{The photon may instead strike the entire hadron coherently.
While suppressed relative to the parton model diagram, it is not as
suppressed as the diagram at large $gN$.}
\end{figure}

Still, for the currents to strike the hadron in this way requires that
the entire hadron shrink down to a size of order $q^{-1}$, a
fluctuation which has amplitude proportional to $q^{-(\Delta-1)}$,
where $\Delta$ is the dimension of the lowest-dimension operator which
can create this hadron.  More precisely, when spin and associated
momentum factors are accounted for, the suppression factor becomes
$q^{-(\tau-1)}$, where $\tau$ is the twist of the lowest-twist
operator which can create the hadron. (A similar factor governs
scattering amplitudes at large angles \cite{PShard}).  Since all such
operators have charge $\QQ$, the dominant contribution to
\Eref{Msschemb} of this type should scale as $(q^2)^{-(\tau_\QQ-1)}$,
where $\tau_\QQ$ is the minimum twist of operators in the
theory with charge $\QQ$.  Indeed, the second term in \Eref{Msschemb}
has this form.

At finite $N$, hadrons are interacting, and therefore hadron number is
not conserved.  Any low-lying eigenstate of the Hamiltonian is an
admixture of a single hadron with a small admixture (of order
$1/N^{k-1}$) of a $k$-hadron state.  More physically, the parent
hadron surrounds itself with a diffuse cloud of other hadrons, some of
them uncharged but some charged.  What is the probability that the
currents will strike not the parent hadron but an electrically charged
hadron in the cloud?  Clearly it must vanish as $N\to\infty$; but the
momentum suppression factor for this process is $(q^2)^{-(\tau_{\rm
c}-1)}$, where $\tau_{\rm c}$ is the lowest-twist operator which can
create a charged object in the cloud.  The cloud-scattering
contribution will be much less suppressed at large $q^2$ than the
parent-scattering, unless $\tau_\QQ=\tau_{\rm c}$.  The third term in
\Eref{Msschemb}, then,
represents scattering of the current off the diffuse cloud
surrounding the parent hadron.
The product $C_{n,p}^{(s)}a_{n,p}$ is related, in analogy
to the parton case, to a {\it hadron} distribution function
in the cloud.

\begin{figure}[h]
\begin{center}
\leavevmode
\epsfbox{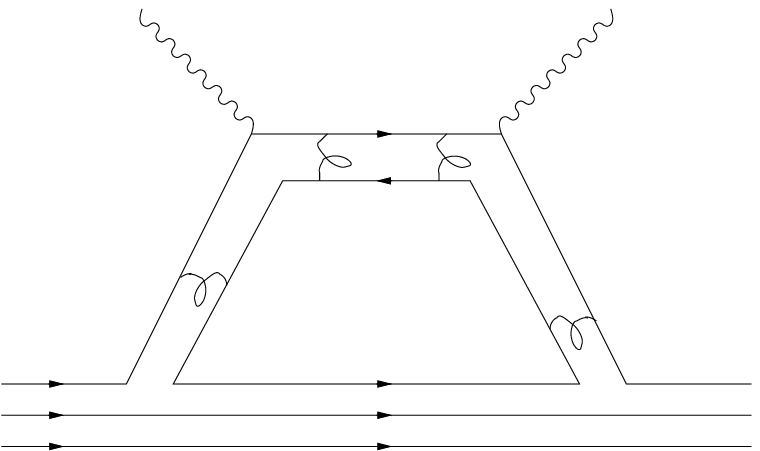}
\end{center}
\caption{The parent hadron has a cloud of smaller hadrons which
surrounds it. At very large $q^2$ the photon is most likely to 
strike the smaller hadron.}
\end{figure}

\section{Strong coupling: string theory analysis}

\subsection{Introduction}

At large 't Hooft parameter, the gauge theories of interest have a dual string
description, in which we can calculate the functions $F_{1,2}$ directly.
In this section we carry out this calculation, verifying the conclusions of
the OPE analysis as well as obtaining more detailed information about the
$x$-dependence.

For conformal gauge theories, the dual string theory lives in a space $AdS_5
\times W$, where $W$ is $S^5$ or other Einstein space.  The metric is
\bel{metric}
ds^2 = \frac{r^2}{R^2} \eta_{\mu\nu} dy^\mu dy^\nu + \frac{R^2}{r^2}dr^2 + R^2
\widehat {ds}_W^2\ ,
\ee
where $R = (4\pi gN)^{1/4}
\alpha'^{1/2}$ is the AdS radius.  The four coordinates $y^\mu$ are
identified with those of the gauge theory, while $r$ is the holographic radial
coordinate; coordinates on $W$ will be denoted $\Omega$.  The key feature of
this geometry is the gravitational redshift (warp factor) multiplying the
four-dimensional flat metric.  The conserved momenta, which are identified
with those of the gauge theory, are $p_\mu = -i
\partial_\mu$.  The momenta as seen by a local local inertial observer in ten
dimensions are
\bel{momred}
\tilde p_\mu = \frac{R}{r} p_\mu\ .
\ee
The characteristic energy scale in ten dimensions is $R^{-1}$ (up to powers
of the dimensionless 't Hooft coupling), so the characteristic
four-dimensional energy is
\bel{echar}
E \sim \frac{r}{R^2}\ .
\ee
Thus, four-dimensional energies depend on the five-dimensional position,
going to zero at the horizon $r = 0$, and diverging at the AdS boundary
$r = \infty$.

In confining theories, the geometry is approximately of the above form at
large $r$, but is modified at radii corresponding to the gauge theory mass
gap,
\bel{r0}
r \sim r_0 = \Lambda R^2\ ,
\ee
so that the warp factor no longer goes to zero.
The details of the small-radius geometry, and the form of the
space $W$, depend on the precise gauge theory and on the mechanism by which
conformal invariance is broken.  However, the dynamics of interest for large
$q^2$ takes place at $r \gg r_0$ where the conformal
metric~\eref{metric} holds, and does not depend on the detailed form of $W$.

Qualitatively, such a theory is similar to QCD, which
also has (nearly) conformal physics in the
ultraviolet, and exhibits confinement and a mass gap in the infrared.
In QCD the ultraviolet is nearly Gaussian, and one sees approximate
Bjorken scaling, but any theory in this class will be expected
to exhibit approximate Kogut-Susskind evolution, which includes
Bjorken scaling as a special case.

We will imagine that we are considering a theory with
a $U(1)$ symmetry current, whose associated
charge is carried by light degrees of freedom
which are present inside
hadrons.  We can then
study the current-current matrix element, \Eref{Wdef}, in the
hadron of interest, and extract from that computation
the same information about hadronic structure functions
that would be obtained by deep inelastic scattering via
a photon coupled to the $U(1)$ current.

\subsection{Examples}

The calculation is largely independent of the details
of the gauge theory, but for completeness we briefly discuss some
examples of theories in which this physics can be
concretely studied.  This section lies outside the mainstream
of the paper and may be skipped.

The simplest way to break conformal invariance is to begin with the \nfour\
theory and add
\none\ supersymmetric mass terms for all
superfields except a single \none\ vector
multiplet.  According to the AdS/CFT dictionary~\cite{GKPW}, this
`\none$^*$' theory corresponds to deforming the string theory on $AdS_5 \times
S^5$ by a nonnormalizable three-form flux at the boundary.  This flux is a
perturbation in the conformal regime but its effect becomes large at small
radius.  The resulting geometry was found in ref.~\cite{PS}.  One or more
expanded five-branes appear near $r_0$, depending on the gauge
theory phase; in the
confining phase there is a single NS5-brane.  One can also add mass terms to
break the supersymmetry completely, but the regime of stability of the
resulting theories has not been precisely determined.

In cascading theories, the transverse space $W = T^{1,1}$ is topologically
$S^3 \times S^2$ with appropriate three-form fluxes.  The \none\ gauge theory,
$SU(N + M) \times SU(N)$ with bifundamentals, cascades down to pure \none\
$SU(M)$  (in the simplest case where $M$ divides $N$), which confines.  The
corresponding geometry is smoothly cut off at small $r$~\cite{ikms}.  In this
case the geometry is not precisely conformal at large $r$, but evolves
logarithmically due to the cascade; we will ignore this slow evolution.

In both of these examples there are continuous symmetries of $W$,
respectively $SO(3)$ and $SU(2) \times SU(2)$, which correspond
to global symmetries of the gauge theory.  We will consider DIS via the
corresponding currents.  We should point out that neither of these examples
exhibits the transition in the precise form discussed in the previous
section, for the simple reason that in the limit of small $gN$ everything
decouples except for a pure
\none\ gauge theory which has no continuous 
global symmetries --- all fields transforming
under the global symmetries of $W$ have masses that are exponentially large
compared to the confinement scale.
However, this is not relevant, for
several reasons.  First, our present concern is only to make
contact with the previous discussion of large $gN$, not to follow the
transition to small $gN$.  Second, it does not mean that the
transition does not occur --- it means only that there are no spin-one
currents we can use to observe it.  One should remember that there is
nothing sacred in spin-one currents (unless one assigns religious
significance to light, for which, admittedly, there is precedent)
and our extended discussion of such currents reflects the specific fact
that in nature we have no other option for probing QCD.  At
a purely theoretical level, we can imagine using spin-2 gravitons,
corresponding to the operator product $T_{\mu\nu} T_{\sigma\rho}$; the earlier
analysis generalizes directly and this gravitational DIS exhibits the
transition.  Third, there are various generalizations that would give weakly
coupled theories with nontrivial global symmetries: keeping one of the masses
zero in
\none$^*$ theory; orbifolding the \none$^*$ theory~\cite{PS};
adding whole D3-branes to the cascading theories; or, adding D7-branes to
either theory.  In each of these examples the details of the decoupling are
somewhat intricate, and we will not go into this subject here.

\subsection{Computation at finite $x$}

We will now use the dual string theory to calculate the matrix element
\bel
{Wagain}
{\sf T}^{\mu\nu} = i \bra{P,\QQ} T(\tilde J^\mu(q)J^\nu(0)) \ket{P,\QQ}\ ,
\ee
(where $\tilde J^\mu$ denotes the Fourier transform),
or at least its imaginary part
\begin{eqnarray}
{\rm Im}\,{\sf T}^{\mu\nu} &=& \pi \sum_{{P}_X,X} \bra{P,\QQ}
J^\nu(0)
\ket{P_X,X}\,
\bra{P_X,X} \tilde J^\mu(q) \ket{P,\QQ}\nonumber\\
&=&
2 \pi^2 \sum_{X} \delta(M_X^2 + [P{+}q]^2)\,
\bra{P,\QQ} J^\nu(0) \ket{P{+}q,X}\,
\bra{P{+}q,X} J^\mu(0) \ket{P,\QQ}\ ,
\label{wim}
\end{eqnarray}
which is what appears in DIS.

We will use indices $M,N,\ldots$ to denote all ten spacetime dimensions,
separating into $m,n,\ldots$ on $AdS_5$ and $a,b,\ldots$ on $W$; the former
separate further into $(\mu,r)$.   We must be a bit careful to distinguish
the flat four-dimensional metric from the warped ten-dimensional metric.  The
momenta
$P_\mu$, $q_\mu$, the polarization
$n_\mu$, and the currents $J_\mu$ will be regarded as four-dimensional
quantities and will be raised or contracted with $\eta^{\mu\nu}$.  Indices
$M,N,\ldots$, $m,n,\ldots$, and $a,b,\ldots$ will be raised or
contracted with the ten-dimensional metric.  An invariant written
without a tilde refers to the four-dimensional gauge theory kinematics, e.g.
$t = -q^2 = -\eta^{\mu\nu} q_\mu q_\nu$.  An invariant with a tilde, e.g.
$\tilde t$, refers to the kinematics in the ten-dimensional metric.

The initial/final hadron is dual to a
string state with wavefunction
\bel{wfun}
\Phi = e^{i P \cdot y} \psi(r,\Omega)
\ee
in ten dimensions.
The function~$\psi(r,\Omega)$ is a
{\it normalizable mode}~\cite{GKPW} in the cut-off $AdS_5 \times W$ space,
like a cavity mode in a box.  For a
given hadron this function will be an eigenstate of the appropriate
Laplacian.  Different eigenfunctions correspond from the four-dimensional
point of view to different hadrons, so that the full hadron spectrum is given
by summing over the different string states and over the different radial Kaluza-Klein 
(KK) modes of each.  (The term ``radial KK modes'' refers to four-dimensional modes reduced 
both on $W$
{\it and} the $AdS$-radial direction; for five-dimensional modes reduced 
on $W$ only, we will use simply ``KK modes''.)
We will take the initial and final hadron to be
unexcited strings, massless in ten dimensions, and for simplicity will focus
on the spinless dilaton.  The currents correspond to perturbations of the
boundary conditions at
$r = \infty$, which excite {\it nonnormalizable modes} in the
bulk~\cite{GKPW}.  Schematically, the calculation is then as shown in
Fig.~2.
\begin{figure}[t]
\begin{center}
\leavevmode
\epsfbox{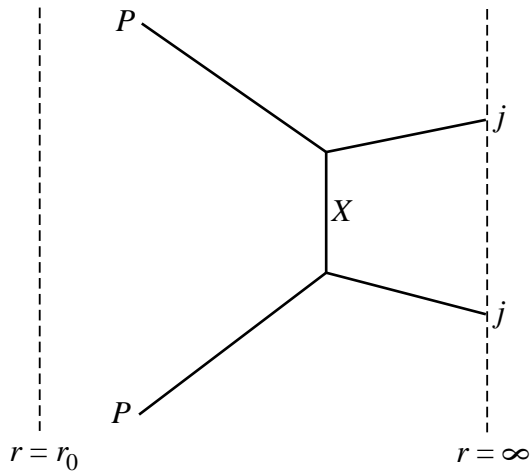}
\end{center}
\caption{AdS/CFT prescription for ${\sf T}^{\mu\nu}$.  The modification of
AdS space at small radius is shown schematically as a sharp cutoff.}
\end{figure}

For reasons that will soon become apparent, the bulk interaction occurs in
the large-radius region, $r \gg r_0$, where the spacetime is essentially a
product $AdS_5 \times W$.  The current that couples to the hadron 
corresponds to
an isometry of $W$, with Killing vector $v_a$.\footnote
{One could consider DIS coupling to other currents, such as the Chan-Paton
symmetries of D7-branes, or even to local operators that are not
conserved currents.  The treatment would be similar, in that the local operator
would excite some nonnormalizable mode.}
It excites a
nonnormalizable mode of a Kaluza-Klein gauge field,
\bel{kka}
\delta g_{m a} = A_m(y,r) v_a(\Omega)\ .
\ee
The boundary limit of $A_\mu$ is the external
potential in the gauge theory:
\bel{lima}
\lim_{r \to \infty} A_\mu(y,r) = A_\mu(y)|_{\rm 4d}\ .
\ee
There is no normalization factor to worry about: a state of unit charge
in the gauge theory maps to a state of unit charge in the bulk.  A
nonnormalizable perturbation with boundary condition
\bel{bound}
A_\mu(y,\infty) = n_\mu e^{i q \cdot y}
\ee
then corresponds to the operator insertion $n_\mu \tilde J^\mu(q)$.  The gauge
field satisfies Maxwell's equation in the bulk, $D_m F^{mn} = 0$.  It is
convenient to work in the Lorenz-like gauge
\bel{lorenz}
i \eta^{\mu\nu} q_\mu A_\nu + R^{-4} r \partial_r (r^3 A_r) = 0\ .
\end{equation}
The field equations are then
\begin{eqnarray}
- q^2 A_\mu + R^{-4} r \partial_r( r^3 \partial_r A_\mu ) &=& 0 \
,\nonumber\\
- q^2 A_r + R^{-4} \partial_r \{ r \partial_r (r^3 A_r) \} &=& 0 \ .
\end{eqnarray}
The solution, with the given boundary and gauge conditions, is
\begin{eqnarray}
A_\mu &=& n_\mu e^{iq\cdot y} \frac{qR^2}{r} K_1(qR^2/r)  \ ,\nonumber\\
A_r &=&  i q \cdot n e^{iq\cdot y} \frac{ R^4}{r^3}
K_0(qR^2/r)\ .
\end{eqnarray}
(where by $q$ alone we mean $\sqrt{q^2}$).

Note that the potential falls off rapidly at $r < qR^2$, from the
exponential behavior of the Bessel functions: the further $q$ is from the
mass shell, the less the perturbation propagates into the AdS interior.  We
are interested in hard scattering, $q \gg \Lambda$, and so
the interaction must occur in the large-$r$ conformal region,
\bel{intrad}
r_{\rm int}\sim qR^2 \gg  r_0 = \Lambda R^2\ .
\ee

We can then use the leading behavior for the wavefunction of the initial and
final hadron, as in Ref.~\cite{PS},
\bel{wfasym}
\Phi_{\rm i} = \Phi_{\rm f}^* \equiv \bra{0} \Phi \ket{P,\QQ}
\sim e^{i P \cdot y} f(r) Y(\Omega)\ ,\quad f(r) \propto
r^{-\Delta}\ .
\ee
In a conformal theory, the mass eigenstates would be of this precise form,
with $\Delta$ the conformal dimension of the state, which is also the
dimension of the local operator $\KK$ that creates the state from vacuum.
In our case, the nonconformal dynamics at small $r$ causes a mixing between
such terms, and the term with smallest $\Delta$ dominates at large $r$.
The normalization of this state was explained in the appendix to
Ref.~\cite{PS}: for a canonically normalized scalar field, canonical
quantization gives
\begin{equation}
\int dr\,d^5\Omega \,e^{2A} \sqrt{g_\perp}\, |\psi(r,\Omega)|^2 = 1
\ , \label{normint}
\end{equation}
where
\bel{foursix}
ds^2 = e^{2A(r,\Omega)} \eta_{\mu\nu} dy^\mu dy^\nu + ds^2_\perp\ .
\ee
The normalization integral is dominated by the IR region $r\sim r_0$, where
$e^{2A}  \sqrt{g_\perp} \sim r_0 R^4$.  It follows that $\psi(r_0,\Omega)
\sim 1/R^2 r_0$ and
\bel{fnorm}
f(r) = \frac{c_{\rm i}}{R^4 \Lambda} (r/r_0)^{-\Delta}
\ee
with $c_{\rm i}$ a dimensionless constant.  We have defined the angular
wavefunction to have the dimensionless normalization
\bel{angnorm}
\int d^5 \Omega \,\sqrt{\hat g_W}\, |Y(\Omega)|^2 = 1\ ,
\ee
where $(\hat g_W)_{ab}$ is the dimensionless metric as defined in
\eref{metric}.  The details of conformal
symmetry breaking enter only through the value of $c_{\rm i}$.

Now let us consider the nature of the intermediate state $X$.  Since we are
working in the leading large-$N$ approximation, only single-hadron
(= single-string) states will contribute.  The important issue is whether
these are massless or excited strings.  In the gauge theory,
\bel{mx2}
s = -(P + q)^2 = q^2 \biggl( \frac{1}{x} - 1 \biggr)\ .
\ee
With the red shift~\eref{momred}, the corresponding ten-dimensional scale is
\begin{eqnarray}
\tilde s = -g^{MN}P_{X,M} P_{X,N} &&\leq
-g^{\mu\nu}(P+q)_\mu (P+q)_\nu
\nonumber\\
&&\stackrel{<}{\sim}
\frac{R^2}{r_{\rm int}^2} q^2 \biggl( \frac{1}{x} - 1
\biggr) = \frac{1}{\alpha' (4\pi gN)^{1/2}} \biggl( \frac{1}{x} - 1
\biggr)\ .\label{tildemx2}
\end{eqnarray}
The 't Hooft parameter appears in the denominator, so if
$(gN)^{-1/2}\ll x<1$ we have $\alpha'\tilde s\ll1$.  Thus for moderate
$x$ only massless string states are produced, and we are dealing
with a supergravity process.  The case that $x$ is small enough to
allow excited strings will be taken up in the next section.

The relevant supergravity interaction, inserting the metric
perturbation~\eref{kka}, is
\bel{mincoup}
\int d^{10} x \sqrt{-g}\, A^m v^a \partial_m \Phi \partial_a \Phi\ .
\ee
The intermediate state is again a dilaton --- there is no mixing in
this case.  We will take the dilaton to be in a charge eigenstate,
\bel{charge}
v^a \partial_a Y(\Omega) = i\QQ Y(\Omega)\ .
\ee
The matrix element of the interaction reduces to the minimal coupling
\bel{mincoup2}
  i\QQ \int d^{10} x \sqrt{-g}\, A^m (
  \Phi_{\rm i} \partial_m \Phi^*_X - \Phi^*_X \partial_m  \Phi_{\rm i})\ ,
\ee
and similarly for the final interaction vertex.
This is equal to the gauge theory matrix element
\bel{gmatel}
n_\mu \bra{P_X,X} \tilde J^\mu(q) \ket{P,\QQ} = (2\pi)^4 \delta^4(P_X {-} P
{-} q) n_\mu \bra{P{+}q,X} J^\mu(0) \ket{P,\QQ} \ .
\ee

In the AdS region the wavefunction $\Phi_X$ satisfies the five-dimensional
Klein-Gordon equation for a scalar of $M^2 = \Delta(\Delta - 4)/R^2$, with
solution
\begin{equation}
\Phi_X = e^{i (P+q) \cdot y} \frac{C}{r^2}  J_{\Delta-2} (s^{1/2} R^2/r)
Y(\Omega)\ . \label{phix}
\end{equation}
Note that $X$ will be a high radial KK excitation, as the
mass-squared~\eref{mx2} grows with $q^2$.  Correspondingly, the turning radius
of the Bessel function, $s^{1/2} R^2$, is large compared to $r_0$, and we
must use the full form of the Bessel function rather than the asymptotic
behavior~\eref{wfasym} used in the external states.  The normalization
integral~\eref{normint} is again dominated by
$r \sim r_0$, giving
$C = c_X s^{1/4} \Lambda^{1/2}$ with $c_X$ another dimensionless
constant.

We can now assemble all factors to obtain
\begin{eqnarray}
n_\mu \bra{P{+}q, X}  J^\mu(0) \ket{P,\QQ} &=& 2 \QQ c_{\rm i} c_X
s^{1/4}
\Lambda^{\Delta-1/2} q\, n_\mu \biggl(P^\mu + \frac{q^\mu}{2x}
\biggr) \nonumber\\
&& \qquad\qquad\qquad \times \int_0^{1/\Lambda} dz\, z^{\Delta}
J_{\Delta-2} (s^{1/2} z) K_1(qz)
\label{matfin}
\end{eqnarray}
where $z = R^2/r$.  The $n_\mu q^\mu$ term comes in part from $A_r$; the
latter must be rewritten using the Bessel recursion relation to obtain
the form~\eref{matfin}, but the result is guaranteed by gauge invariance.
The upper limit of integration is essentially $\infty$, and the integral is
then
\begin{equation}
2^{\Delta-1} \Gamma(\Delta) \frac{s^{\Delta/2-1} q}{(s +
q^2)^\Delta}\ ,
\end{equation}
giving
\begin{eqnarray}
\bra{P{+}q, X}  J^\mu(0) \ket{P,\QQ} = 2^\Delta \QQ c_{\rm i} c_X
\Gamma(\Delta)
\biggl(P^\mu + \frac{q^\mu}{2x}
\biggr)
\Lambda^{\Delta-1/2} \frac{s^{\Delta/2-3/4} q^2}{(s +
q^2)^\Delta}\ .
\label{matfin2}
\end{eqnarray}

To find the imaginary part~\eref{wim} it remains to square the above result
and sum over radial excitations.  We can estimate the density of states by a
hard cutoff at a radius $r_0$, so that the spacing of the zeros of the Bessel
function~\eref{phix} gives
\bel{spacing}
M_{n} = n\pi\Lambda\ .
\ee
In the leading large-$N$ approximation the structure functions are necessarily
a sum of delta functions, but at large $q$ their spacing is close and so
\bel{density}
\sum_{n} \delta(M_n^2 - s) \sim (\partial M_n^2/\partial n)^{-1} \sim
(2\pi s^{1/2}\Lambda)^{-1}\ .
\ee

Assembling all factors gives the final result
\bel{tfin}
{\rm Im}\,{\sf T}^{\mu\nu} = A_0 \QQ^2
\biggl(P^\mu + \frac{q^\mu}{2x} \biggr) \biggl(P^\nu + \frac{q^\nu}{2x}
\biggr)
\Lambda^{2\Delta - 2} q^{-2\Delta} x^{\Delta+2} (1-x)^{\Delta - 2}
\ .
\ee
The entire result is fixed up to the IR-dependent normalization constant
$A_0 = 2^{2\Delta} \pi |c_{\rm i}|^2 |c_X|^2 \Gamma(\Delta)^2$.  In 
terms of the
structure functions~\eref{ftf}, this is
\bel{ffin}
F_1 = 0\ ,\quad F_2 = \pi A_0 \QQ^2 ( {\Lambda^2}/{q^2} )^{\Delta
- 1} x^{\Delta+1} (1-x)^{\Delta - 2}\ .
\ee

\subsection{Extension to spin-$\frac{1}{2}$}

We can readily extend this to spin-$\frac{1}{2}$ hadrons, corresponding to
supergravity modes of the dilatino.  In the conformal region the dilatino field
separates
\bel{dino}
\lambda = \psi(y,r) \otimes \eta(\Omega)\ ,
\ee
where $\psi(r)$ is an $SO(4,1)$ spinor on $AdS_5$ and $\eta(\Omega)$ is an
$SO(5)$ spinor on $W$.  For $\eta$ an appropriate eigenfunction the field
equation reduces to a five-dimensional Dirac equation
\bel{dirac}
-i D\hspace{-7.7pt}\slash \hspace{1.1pt}\psi = m \psi\ .
\ee
The solution to this is~\cite{dirsol}
\bel{dirsol}
\psi = e^{i p \cdot y} \frac{C'}{r^{5/2}} \Bigl[
J_{mR - 1/2}(\mu R^2/r) P_+ + J_{mR + 1/2}(\mu R^2/r) P_- \Bigr] u_\sigma\ ,
\ee
where
\bel{defu}
p\hspace{-5.3pt}\slash u_\sigma = \mu u_\sigma\quad (\sigma = 1,2)\ ,\quad
\mu^2 = -p^2\ ,\quad P_{\pm} = \frac{1}{2} ( 1 \pm \gamma^{\hat r} )\ .
\ee
Note that $\hat r$ is a tangent space index, and that $\gamma^{\hat r}$ is the
same as the four-dimensional chirality $\gamma_5$.

For the initial hadron,
$\mu r \ll 1$ in the interaction region and
\bel{psiinit}
\psi_{\rm i} \approx e^{i P \cdot y} \frac{c'_{\rm i}}{\Lambda^{3/2} R^{9/2} }
(r/r_0)^{- mR - 2} P_+ u_{\rm i\sigma} \ .
\ee
 From the $r$-dependence, we identify the conformal dimension of the state as
\bel{psidim}
\Delta = mR + 2\ ;
\ee
the spinor index does not affect the scaling because it is an inertial (tangent
space) index.  For the intermediate hadron, $\mu = s^{1/2} \gg \Lambda$ and
\bel{psix}
\psi_{X} \approx e^{i (P+q) \cdot y} \frac{c'_{X} s ^{1/2} \Lambda^{1/2}
R^{1/2}}{r^{5/2}}
\Bigl[
J_{mR - 1/2}(\mu R^2/r) P_+ + J_{mR + 1/2}(\mu R^2/r) P_- \Bigr] 
u_{X\sigma} \ .
\ee
We have normalized
\bel{spinnorm}
\int d^5 \Omega \,\sqrt{\hat g_W}\, \OL{\eta}(\Omega) \eta(\Omega) = 1\ .
\ee

Let us take the polarization $n_\mu$ to be orthogonal to $q^\mu$ so that $A_r =
0$; this is sufficient to read off the two structure functions.  Then
\begin{eqnarray}
n_\mu \bra{P_X,X,\sigma'}  J^\mu(0) \ket{P,\QQ,\sigma} &=&  i\QQ \int d^{6}
x_\perp \sqrt{-g}\, A_m \OL{\lambda}_X \gamma^m \lambda_{\rm i} 
\label{fercur}\\
&=& i\QQ c'_{\rm i} c'_X
s^{1/4}
\Lambda^{\tau-1/2} q\, n_\mu\, \OL{u}_{X\sigma'} \gamma^{\hat \mu} u_{\rm i
\sigma}
\nonumber\\
&&\qquad\qquad \times \int_0^{1/\Lambda} dz\, z^{\tau}
J_{\tau-2} (s^{1/2} z) K_1(qz)\ . \label{ferint}
\end{eqnarray}
We have written the result in terms of $\tau = \Delta - 1/2$, in terms of
which it is nearly identical to the bosonic matrix element~\eref{matfin}
(where $\tau =
\Delta$).  Summing over radial excitations and final state spin, and averaging
over initial spin, then gives
\bel{tferm}
n_\mu n_\nu{\rm Im}\,{\sf T}^{\mu\nu} = A'_0 \QQ^2\,
\frac{1}{4} {\rm Tr}\Bigl[P\hspace{-7.3pt}\slash \,n\hspace{-6pt}\slash
(P\hspace{-7.3pt}\slash + q\hspace{-6pt}\slash + s^{1/2}) n\hspace{-6pt}\slash
  P_+ \Bigr]
\Lambda^{2\tau - 2} q^{-2\tau} x^{\tau+2} (1-x)^{\tau - 2}
\ .
\ee
The polarization trace is
\bel{poltra}
\frac{1}{4}  {\rm Tr}\Bigl[P\hspace{-7.3pt}\slash \,n\hspace{-6pt}\slash
(P\hspace{-7.3pt}\slash + q\hspace{-6pt}\slash + s^{1/2}) n\hspace{-6pt}\slash
  P_+ \Bigr] = (P \cdot n)^2 - \frac{1}{2} P \cdot q\, n^2\ .
\ee
Then
\bel{fermfs}
F_2 = 2F_1 = \pi A'_0 \QQ^2 ( {\Lambda^2}/{q^2} )^{\tau
- 1} x^{\tau+1} (1-x)^{\tau - 2}\ .
\ee

\subsection{Discussion}

The $q^2$-dependence of the results~\eref{ffin}, \eref{fermfs} is precisely as
deduced from the OPE for the leading large-$N$ behavior at strong coupling.  It
corresponds to the second term in \Eref{Msschemb}, coming from the double-trace
operators of twist $2\tau_p$.  As we described above, it represents 
scattering of
the electron off the entire hadron.

It is interesting also to understand this
$q^2$ dependence in the string picture.
A naive interpretation of string theory would suggest a very soft
amplitude, falling exponentially in $q^2$, because there are no partons in
the string.  It is the
warped geometry that permits power law scattering.\footnote
{Indeed, it is the warped geometry that permits the introduction of local
currents in the first place~\cite{size}.}
This can be understood pictorially, from
Fig.~3.  The string tension $\tilde T$ in an inertial frame is constant, but as
measured in the $x^\mu$ coordinates it grows with $r$ due to the warp factor,
$T = r^2 \tilde T /R^2$.  Correspondingly, the characteristic size of a
string in an inertial frame is constant, but its projection on the $x^\mu$
coordinates is smaller at larger $r$.  The most 
efficient way for
a string to undergo hard scattering is to tunnel to large enough $r$ that its
size is of order the inverse momentum transfer.  This costs only a power
law suppression, from the conformal dimension of the state.
\begin{figure}
\begin{center}
\leavevmode
\epsfbox{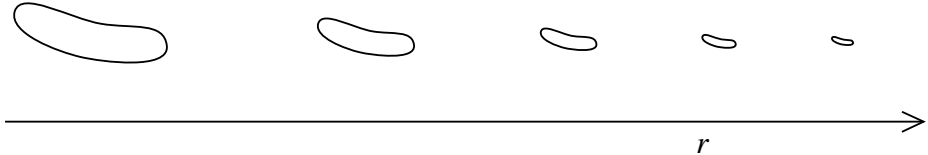}
\end{center}
\caption{An optical illusion: all the strings are the same size.  Strings of
constant size in an inertial frame, at different radial positions $r$, have
different coordinate sizes.}
\end{figure}

This power-law effect is true both for the elastic scattering studied in ref.~\cite{PShard} and
for the inelastic scattering considered here.  However, in the elastic case 
the scaling in
the string limit is qualitatively the same as in the parton model.   The reason is that in both the 
parton and string limits, elastic scattering at large momentum transfer requires the 
entire hadron to shrink to small size, and so is determined by the overall dimensional
scaling of the hadron wave function.   In both cases the magnitude of the wavefunction in this region is
determined simply by the conformal dimension of the gauge-invariant operators
which can create the state.\footnote{The possibility
of the Landshoff process --- multiple parton-parton scattering ---
in the parton case does make the story potentially more complicated,
but we will not discuss this here; we expect this is absent in the string limit.}
For hadrons that mix
with chiral states of the superconformal algebra, this dimension is the same
at large and small 't Hooft parameter; more generally the powers involved
are of order 1 but are never zero.

However, in {\it inelastic} scattering it is not necessary for the entire hadron to shrink to small size.
The photon may strike only a fraction of the hadron, leading to different dimensional scaling.
In the parton limit, the photon may strike as little as
a single parton.  The probability that one parton will be small while the rest of the hadron is large
is clearly greater than the probability that the entire hadron is small, or even that two partons shrink
together to a small size; thus one-parton scattering
dominates.  Since the operators which create and destroy a single parton can have twist 2 in this limit,
they lead to Bjorken scaling.  By contrast, in the string limit, as evident from Fig.~3, the probability
that a small fraction of the string will tunnel to large $r$ while the rest remains at small $r$ is highly
suppressed.  This implies the four-dimensional hadron  
does not contain pointlike partons at large 't Hooft parameter.  
This corresponds to the fact that operators which couple to one or more partons in
a gauge-non-invariant combination have large twist.  The string is 
able to scatter
inelastically only in the same way as it scatters elastically, 
by tunneling to large $r$ where its projection onto $x^\mu$ is
small, and then scattering as a unit.  As in the elastic case,
the magnitude of the wavefunction in this region is
determined simply by the conformal dimension of the state. 
The entire hadron must shrink to a size 
of order $q^{-1}$; our calculation
shows that the probability of this is
$q^{-2(\tau_p - 1)}$.  Since the gauge-invariant operators which can create
a hadron in \nonestar\ and many similar theories have $\tau_p\geq 2$, 
we never recover
Bjorken scaling in the string limit.  (Note that formally a hadron with $\tau_p = 1$
behaves as a single pointlike parton, as in other contexts.)

In the leading large-$N$ limit we are studying the internal
dynamics of a single string, with string production turned off.  From the OPE
analysis we concluded that at finite $N$ a subleading piece would ultimately
dominate at large $q^2$.  The corresponding supergravity process is depicted
in Fig.~4.  The incoming hadron splits into two, one of which (B) has the
minimum twist $\tau_{\rm c}$ of any charged hadron.  (In \nonestar, for
example, the lowest-twist charged hadron is a $\Delta=2$ state of the
dilaton.)  Hadron B then tunnels to large $r$ and interacts with the current.
The interaction of B with the the external current should be essentially the
same as above with
$\tau_{\rm c}$ in place of $\Delta$.  The splitting of the initial hadron
into A and B occurs in the IR region and so depends on details of the model.
This can presumably be encoded into a distribution function for B to carry a
fraction $x$ of the incoming energy, which we simply need to convolve 
with the above
result at $\tau = \tau_{\rm c}$.  We have not attempted
to calculate the distribution function, but we believe it can be done 
within the
supergravity approximation.
\begin{figure}
\begin{center}
\leavevmode
\epsfbox{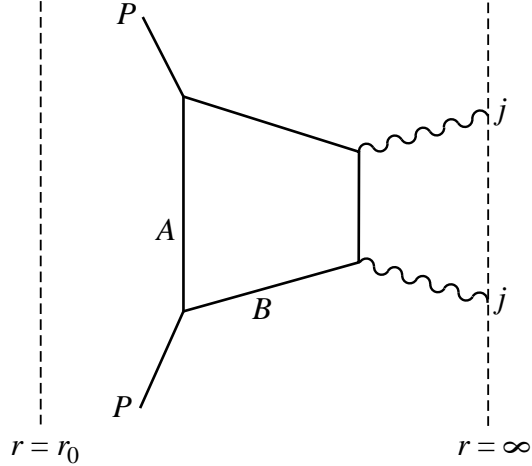}
\end{center}
\caption{Subleading contribution in $N$.  The incoming hadrons splits into
hadrons A and B, with B having minimum twist $\tau_{\rm c}$.}
\end{figure}

The $q^2$ dependence thus reproduces what we deduced from the
OPE.  We now consider the remaining details of the string result.  First, we
see that the function $F_2$ is the same for both
the dilaton and dilatino.  This is reasonable, since (as can be seen by
expanding the propagator of the scattered particle in powers of the
external photon) the function $F_2$ 
generally measures spin-independent
information.  By contrast, $F_1$ is proportional to the Casimir of the
scattered object under the Lorentz group, so it vanishes for the scalar hadron,
as it also would for a scalar parton.  The distinction between spinor parton and
spinor hadron scattering (aside from the $q^2$ dependence) shows up 
in the modification of the Callan-Gross relation.
A scattered parton in the parton model has momentum $xP$, while in our case
the scattered hadron has momentum $P$.  This missing factor of $x$ is the same
one missing from the Callan-Gross relation in our computation \eref{fermfs}.

We do not have a physical interpretation of the full $x$-dependence of our
result, but we can
understand the behavior near $x = 1$.  For $x \sim 1$ the 
mass-squared of the
intermediate state is
\bel{mxsim1}
s  = q^2 (1-x)\ .
\ee
If we take $q^2 \to \infty$ with $s$ held fixed (so that $1 - x \to 0$),
then the matrix element $\bra{P_X,X} \tilde J^\mu(q) \ket{P,\QQ}$ is
essentially a hard form factor, and is governed by the same kind of scaling
as the elastic amplitudes~\cite{PShard,size}. In this limit, the 
Bessel function
in the wavefunction for $X$ becomes a power law, just as for the initial
hadron.  This reflects the fact that in this kinematic regime both the initial
hadron and the hadron $X$ have to shrink down to size
$q^{-1}$ in order to scatter from the photon.  It follows that the matrix
element is determined by the conformal weight, falling as $q^{2 - 2\tau_p}$,
and so the structure
function is
\bel{f2scale}
F_2 = q^{6-4\tau_p} G_2(q^2 [1-x] )
\ee
for some function $G_2$. But we also know from the
OPE that $F_2$ scales as
$q^{2 - 2\tau_p}$, so it must be that $G_2 \propto (q^2 [1-x] )^{\tau_p -
2}$ which is indeed the behavior found as $x \to 1$.  This same 
argument applies
at small 't Hooft parameter (where $\tau_p$ is the number
$n$ of partons in the initial hadron), but at the final step we require
instead that
$F_2$ be independent of $q^2$, giving $G_2 \propto (q^2 [1-x] )^{2 n -
3}$ and $F_2 \propto (1-x)^{2n - 3}$, up to the usual perturbative scaling
violations~\cite{xtoone}.

\section{The Regge Region}

Our analysis is incomplete in an important respect.
We have noted that the energy-momentum tensor has no anomalous
dimension, and therefore the moments
\bel{Rmom}
M^{(1)}_2 = \int_0^1 dx\,x F_1(x,q^2)\ ,\quad
M^{(2)}_2 = \frac{1}{2} \int_0^1 dx\, F_2(x,q^2)
\ee
have nonzero limits as $q^2 \to \infty$, determined by the $J^\mu J^\nu
T^{\rho\sigma}$ operator product coefficient.  This is not the case for our
result~\eref{ffin}, which gives moments that falls as
$q^{2 - 2\Delta}$.  Such a falloff is correct for the higher moments, with
additional powers of
$x$, so it must be that we have missed a component of $F_{1,2}$ that 
is narrowly
peaked around $x = 0$.

The existence of such a component is also suggested by a more physical
argument~\cite{kosu}.  At small coupling, interactions cause partons to split
and so the structure functions evolve toward smaller $x$.
As the interactions become strong, this evolution becomes more rapid. 
With very
strong interactions, the parton language becomes inapplicable, but one might
still expect a rapid evolution toward small $x$, leaving most of the weight of
the $F_i$ at near $x=0$.  The moments~\eref{Rmom} must be conserved 
by this flow.

The calculation in the previous section is valid as long
as the supergravity approximation holds.  However,
for sufficiently small $x$, this will no longer be the
case.  \Eref{tildemx2} shows that the scattering energy from the
point of view of the local observer in the bulk becomes
of order the string mass scale when $x\sim (gN)^{-1/2}$.
Beyond this point, one must take seriously the fact
that the bulk theory is a string theory.  Since
DIS is extracted from a forward scattering amplitude, it is the
Regge physics of the string which is important here.

In fact this point is used in standard QCD to argue that 
the small-$x$ behavior
of the gluon structure function is dominated by Pomeron
exchange \cite{Abarbanel,FR}.
This is mainly kinematics: the center-of-mass energy-squared
of the photon-hadron system in the DIS process is
$
s  \approx q^2/x
$
as $x\to 0$, so the small $x$ region is the large $s$ region.
The Pomeron --- a trajectory of conjectured glueball states of
increasingly higher spin, analogous to the graviton trajectory
of string theory --- gives a contribution to parton distribution
functions of the form
$f(x,q^2)\sim x^{-\alpha_0}$, where $\alpha_0$ is
the intercept of the Pomeron trajectory with the spin axis in
the $M^2$ {\it vs.}~$J$ Regge plot.  The variation with $q^2$ is
not specified by general arguments,
but it has been argued \cite{Abarbanel} that
in QCD it should be slow.  Since Pomeron exchange also predicts that
the total proton-proton cross-section should grow as $s^{\alpha_0}$,
the intercept can be extracted from data (giving something of order $1.1$)
or predicted from QCD (giving something of order $1.4$.)  The interpretation
of these discrepant values is controversial.
For our purposes it suffices that $\alpha_0$ in QCD is closer
to 1 than  to 2.
That the gluon structure function grows faster than $x^{-1}$
but quite a bit slower than $x^{-2}$ is in fact observed at HERA.

In \nonestar, one could in principle compute the Pomeron
intercept at small $gN$ using the perturbative techniques outlined in
\cite{FR}; we have not done this.  The large $gN$ region is much easier,
however. The Pomeron trajectory {\it is} the graviton trajectory; more
precisely, it is the trajectory whose lowest state is the lowest-lying
spin-two glueball, given as the lowest mode of the graviton inside the
cutoff five-dimensional $AdS$ space. 
An exactly massless spin-two mode would give an intercept of
exactly two. The IR cutoff on the $AdS_5 \times W$ space produces a mass gap of
order $R^{-2}$ in the bulk, and so lowers the 
intercept of the trajectory by order
$\alpha'/R^2 \sim (gN)^{-1/2}$. The intercept 
is therefore $2-{\rm order}(gN)^{-1/2}$, and so
we therefore expect that the DIS amplitudes will
have the behavior
\bel{regexp}
x F_1 \sim F_2 \propto
x^{-1+ O(gN)^{-1/2}}
\ee
at small $x$.  While we will see that this is 
true, the full story is somewhat more subtle.

An important caveat is that we will consider only the leading effect
in the large-$N$ limit, corresponding to string tree level.  In other
words, we are focusing on the internal structure of a {\it single}
string.  Because of the growth of the amplitude with energy, the
production of multiple strings (and ultimately black hole
formation~\cite{giddings}), corresponding to multi-Pomeron exchange,
shadowing, {\it etc.}, dominates at asymptotically large $s$ (small
$x$) for any finite $N$.  We will not address these phenomena here.
For other approaches to fixed momentum
transfer scattering in AdS/CFT, see Ref.~\cite{eikon}.

\subsection{Calculation at small $x$}

In the supergravity calculation we explicitly summed over intermediate states
$X$, which were radial KK excitations, but in the string case it is simpler to sum
implicitly by taking the imaginary part of the forward four-point amplitude.
The momentum invariants in the inertial frame are of order the string
scale, so we might expect the scattering process to be localized on this
scale.  Since this is small compared to the AdS radius, we can take the
flat spacetime string amplitude and fold it into the AdS wavefunctions.
Actually, we will see that at exponentially small $x$ this locality breaks
down, but for now we will assume it.

The string interaction can be written~\cite{GSW}
\bel{stringamp}
{\cal L}_{\rm eff,string} = K G
\ee
with $K$ a kinematic factor and  
\bel{RG}
G = -\frac{\alpha'^3 \tilde s^2}{64} \prod_{ \xi =  s, t, u}
\frac{\Gamma(-\alpha' \tilde \xi/4)}{\Gamma(1+\alpha' \tilde \xi/4)}\ ;
\ee
the prefactor is included for later convenience.  The tildes are
included on the Mandelstam variables because this flat spacetime
amplitude goes over to the inertial frame quantity in curved
spacetime.  Expanding around $\tilde t=0$, one finds, for $\tilde s$
near the positive real axis,
\bel{RG2}
G =  -\left[{1\over \tilde t} 
+ \pi  \cot\left({\pi \alpha'\tilde s\over 4}\right)\right]
\left[{\alpha'\tilde s\over 4e}\right]^{\alpha'\tilde t/2}
\left[1 + {\rm order}(\alpha'\tilde t)\right]
\ee
(To obtain this expression we have used Stirling's approximation, but
when $\alpha'|\tilde t|\ll1$ the form is valid even for $\alpha'\tilde
s\sim 1$.)  The imaginary part from excited strings is
\bel{Gim}
{\rm Im}_{\rm exc}\,G|_{\tilde t \to 0} = \frac{\pi\alpha'}{4}
\sum_{m=1}^\infty
\delta(m - \alpha' \tilde s /4) \ ({m})^{\alpha'\tilde t/2}\
\ .
\ee
The last factor --- the small-angle Regge behavior of the string
amplitude --- becomes important at ultra-small $x$.  This is because,
as we will see later, the ten-dimensional $\tilde t$ is not quite
zero, even though $t=0$.  Instead, it will turn out that $\alpha'\tilde t\sim
(gN)^{-1/2}.$  This implies that when $\alpha'\tilde
t\log(\alpha'\tilde s) \agt 1$, that is, for $\tilde s \sim 1/x$
exponentially large, of order $e^{\sqrt{gN}}$, we must include this
term.  For $e^{-\sqrt{gN}}\ll x \ll {1\over\sqrt{gN}}$,
however, we can ignore it.

To obtain the imaginary part of the forward amplitude we must evaluate
$K|_{\tilde t = 0} $.
One can do this directly from the definition~\cite{GSW} or by noting from
the expansion~\eref{RG2} that it is the same as the coefficient of the
$t$-channel graviton pole.  One finds
\bel{ktzero}
K|_{\tilde t = 0} = \frac{1}{8} \int d^{10}x\,\Bigl\{
4 v^a v_a \,\partial_m \Phi F^{mn} F_{pn} \partial^p \Phi
- ( \partial^M \Phi \partial_M \Phi\,
v^a v_a + 2 v^a \partial_a \Phi \,
  v^b \partial_b \Phi) F_{mn} F^{mn} \Bigr\} \ ,
\ee
with the same index conventions as in \hbox{Section~IV}.  The dilaton
is canonically normalized and the normalization of the vector
potential is defined by the Kaluza-Klein ansatz~\eref{kka}.  We have
written this as an effective action in position space to facilitate
the lift to curved spacetime.  At small $x$, the first term in the
effective action~\eref{ktzero} dominates, because the index structure
gives two extra factors of $P \cdot q$; thus we will focus on this
term. Notice that the explicit dependence on $\QQ$, through $v^a
\partial_a \Phi_{\rm i} = i \QQ \Phi_{\rm i}$, is subleading --- as in
QCD, the
wee parton cloud is universal and does not depend on the quantum
numbers of the hadron.

Evaluating this effective Lagrangian in the AdS matrix element gives
the contribution of excited strings to the imaginary part of the forward
current-current amplitude.  For $x$ not exponentially small, we drop
the $\tilde s^{\alpha'\tilde t}$ factor and obtain
\begin{eqnarray}
n_\mu n_\nu {\rm Im}_{\rm exc}{\sf T}^{\mu\nu} &=&
(K\,{\rm Im}\,G)|_{\tilde t = 0} \nonumber\\
&=&\frac{\pi\alpha'}{2}
\sum_{m=1}^\infty
\int d^{6}x_\perp\sqrt{-g} \,
v^a v_a \,\partial_m \Phi^*_{\rm i} F^{mn}(-q) \delta(m - \alpha' \tilde
s /4) F_{pn}(q) \partial^p \Phi_{\rm i}\ .
\label{Ramp}
\end{eqnarray}
All contractions are now with the full ten-dimensional metric.  The invariant
$\tilde s$ in principle involves a differential operator acting on 
the fields to
its right,
\bel{kinem}
\alpha' \tilde s = \frac{\alpha' s R^2}{r^2} + \frac{\alpha'}{R^2} (r^2
\partial_r^2 + 5 r \partial_r + \hat \nabla_W^2 )\ .
\ee
Everything is smooth in the radial direction, so $r \partial_r = O(1)$, and the
dimensionless Laplacian on $W$ is similarly assumed to be $O(1)$.  
Consequently, $\alpha' \tilde s = {\alpha' s} (R^2/r^2)$ plus corrections
of order $\alpha'/R^2 \sim (gN)^{-1/2}$, which for
excited strings can be neglected compared to the integer $m$ in the delta
function.  This is consistent with our argument that the effective 
interaction is
local on the AdS scale.  (The case $m=0$ is not covered by this argument,
but we have already calculated it
explicitly, without this assumption, in Section IV.)
The delta function then becomes $\delta(m - \alpha' s R^2/4r^2)$,
and so  just fixes the radial integration.

We can now evaluate the amplitude~\eref{Ramp}.  The components of the field
strength are
\begin{eqnarray}
F_{\mu\nu}(q) = i(q_\mu n_\nu - n_\mu q_\nu) \frac{q R^2}{r} K_1(qR^2/r)
e^{i q\cdot y}\ ,
\nonumber\\
F_{\mu r}(q) = (n_\mu q^2 - q_\mu q \cdot n) \frac{R^4}{r^3} K_0(qR^2/r) e^{i
q\cdot y}\ . \label{fmunu}
\end{eqnarray}
which implies
\begin{eqnarray}\label{ffmurho}
F_{\mu p} F_\nu{}^p &=& \biggl(n_\mu - q_\mu \frac{q\cdot
n}{q^2}\biggr)
\biggl(n_\nu - q_\nu \frac{q\cdot n}{q^2}\biggr)
\frac{w^4}{R^2} [K_0^2(w) + K_1^2(w) ]\nonumber\\
&&\qquad\qquad
+ {q^\mu q^\nu} \biggl(n^2 - \frac{q\cdot n \,
q\cdot n}{q^2} \biggr) \frac{w^4}{q^2 R^2} K_1^2(w)
\end{eqnarray}
where $w = q R^2/r$.
(The contributions of $F_{r p} F_\nu{}^p$ and $F_{r p} F_r{}^p$ are 
negligible).   
Then
\begin{eqnarray}
n_\mu n_\nu {\rm Im}_{\rm exc}{\sf T}^{\mu\nu} &=& \frac{\pi \rho |c_{\rm
i}|^2}{8}\Biggl(\frac{\Lambda^2}{q^2} \Biggr)^{\Delta - 1} \,
\sum_{m=1}^\infty w_m^{2\Delta+2}
\Biggl\{ \frac{1}{x} K^2_1(w_m) \biggl(n^2 - \frac{q \cdot n\,q\cdot 
n}{q^2} \biggr)
\nonumber\\
&&{\qquad\qquad\qquad}
+ \frac{4 x}{q^2}\Bigl[K^2_0(w_m) + K^2_1(w_m)\Bigr] \biggl(P \cdot n 
- \frac{P \cdot
q\,n\cdot q }{ q^2}\biggr)^2 \Biggr\}\ ,
\label{Ramp2}
\end{eqnarray}
where $w_m = qR^2/r_m$ and $r_m = {R (\alpha' s )^{1/2}}/{2  m^{1/2}}$.
The dimensionless constant
$\rho$ is from the angular integral
\bel{Rang}
\int d^5\Omega \sqrt{\hat g_W}\, v^a v_a |Y|^2 = \rho R^2
\ee
(note that $v^a$ and
$\partial_a \psi$ are $R$-independent).  The structure functions are then
\begin{eqnarray}
F_1  &=& \frac{\pi^2 \rho |c_{\rm i}|^2}{  4x } \Biggl(\frac{\Lambda^2}{q^2}
\Biggr)^{\Delta - 1}
\sum_{m=1}^\infty w_m^{2\Delta+2}
  K^2_1(w_m)\ ,\nonumber\\
F_2  &=& \frac{\pi^2\rho |c_{\rm i}|^2}{2} \Biggl (\frac{\Lambda^2}{q^2}
\Biggr)^{\Delta - 1}
\sum_{m=1}^\infty w_m^{2\Delta+2}
\Bigl[K^2_0(w_m) + K^2_1(w_m)\Bigr]
\ . \label{w12}
\end{eqnarray}

The argument of the Bessel functions is $w_m \sim x^{1/2} m^{1/2} /(gN)^{1/4}$,
and so when $x \ll (gN)^{-1/2}$ the sums can be approximated by integrals.  Using
\bel{Rdimcon}
I_{j,n}= \int_0^\infty dw\,w^n K_j^2(w) = 2^{n-2}
\frac{\Gamma(\nu + j)\Gamma(\nu - j)\Gamma(\nu)^2}{\Gamma(2\nu)}\ ,\quad
\nu = \frac{1}{2}(n+1)\ .
\ee
so that $I_{1,n}={n+1\over n-1}I_{0,n}$, one finds
\begin{eqnarray}
F_1  &=&  \frac{1}{x^2}\,
\Biggl(\frac{\Lambda^2}{q^2}
\Biggr)^{\Delta - 1} \frac{\pi^2\rho |c_{\rm i}|^2}{  4 (4\pi gN)^{1/2}}
I_{1,2\Delta+3}\ ,\nonumber\\
F_2  &=& 2x {2\Delta+3\over \Delta+2}F_1 \ .
\label{Rw12}
\end{eqnarray}

\subsection{Exponentially small $x$}

As expected, the inclusion of excited strings leads to a new term in the
structure functions, which dominates at small $x$.  This has an 
interesting effect
on the moments
$M_n^{(1,2)}$.  For $n > 2$, the new contribution converges and gives the same
$q^{2 - 2\Delta}$ dependence as found in the supergravity calculation; this
corresponds again to the double-trace operators found in section~III.
However, the momentum sum $n=2$ now {\it diverges} as $x \to 0$.  Thus we need
to understand how the calculation that we have done breaks down in this limit.

We emphasize again that we consider
only the leading effect in the 
large-$N$ limit, corresponding to the internal 
structure of a single
string.  Since the amplitudes grow with energy and we will now study values of
$x$ that are exponentially small in $(gN)^{1/2}$, the following calculation is the
dominant process only
for $N$ that is exponentially large.  Thus we are looking at an 
extreme region of
parameter space, but one that is conceptually interesting.

Note that the $x$-dependence of the structure functions~\eref{Rw12} agrees to
leading order with the field theory expectation~\eref{regexp}.  However, the
$O(gN)^{-1/2}$ correction to the exponent is crucial to the convergence of the
momentum sum rule.  It appears that we need to go to the next order in the 't
Hooft parameter, which is a daunting task.  However, we will argue that the
crucial correction enters in a simple way.  We will first support this by
heuristic arguments, and after carrying out the improved calculation 
we will give
a more formal justification using the string world-sheet OPE.

We claim that we simply need to keep the full Regge form of the string
amplitude by restoring the Regge factor in \eref{RG2}.  The Regge factor 
is nontrivial because, although the four-dimensional $t$ vanishes by 
definition, the ten-dimensional
$\tilde t$ includes in addition derivatives in the transverse directions, as with
$\tilde s$ in \eref{kinem}.  This
$\tilde t$ is small on the string scale, of order $(gN)^{-1/2}$, but its effect
becomes large when $\tilde s$ is exponentially large in $(gN)^{1/2}$, or
equivalently when $x$ is exponentially small.  This modification 
clearly has the
right form to produce the Regge correction to the $x$-dependence of the structure
function, and we will argue later that it is the only important correction.

Notice that the modification is nonlocal, because it involves a power of the
differential operator $\tilde t$.  This is the logarithmic spreading of strings
in the Regge region~\cite{growth}.  Thus the assumption that the interaction is
approximately local breaks down, though in a way that is fairly simple to
incorporate by replacing $\tilde t$ with a curved spacetime 
Laplacian.  The need
for such nonlocality can be understood in various ways.  For one, local
interactions will always contain a factor of $q^{-2\Delta}$ and so cannot give
$q$-independent moments.  This is because the falloff of the vector potential
requires that the interaction occur at $r\agt
R^2 q$, and so we only pick up the $r^{-\Delta}$ tail of
each wavefunction.  For another, the {\it real} part of the amplitude is
nonlocal, from graviton exchange.  That is, the vector background induces a
metric perturbation at second order, which can propagate to smaller $r$ without
exponential suppression.  By analyticity we expect a similar effect in the
imaginary part.

Our prescription is then to modify \eref{Ramp} by inserting $\tilde s^{\alpha'\tilde t/2}\sim
x^{-\alpha'\tilde t/2}$, averaging over the delta functions (since we are at exponentially
large $m$) and writing $\tilde t$ as a differential operator acting in the $t$-channel:
\bel{ampmod}
n_\mu n_\nu {\rm Im}_{\rm exc}{\sf T}^{\mu\nu}= \frac{\pi\alpha'}{2}
\int d^{6}x_\perp\sqrt{-g} \,
v^a v_a \, F^{mn}(-q) F^p{}_{n}(q)
\, x^{-\alpha' \nabla^2/2} (\partial_m \Phi^*_{\rm i} \partial_p \Phi_{\rm i})
\ .
\ee
In the correction term we have replaced $\alpha' \tilde s$ with $1/x$ 
because the
exponentially large parts of these are the same:
\bel{expol}
\alpha' \tilde s \sim \alpha' s R^2 / r_{\rm int}^2 \sim \alpha' s / 
R^2 q^2 \sim
1/(gN)^{1/2}x\ ;
\ee
they differ by a power of the 't Hooft parameter, which is large but not
exponentially so.

We need to identify the Laplacian
that appears.  The dominant components of $\partial_m \Phi^*_{\rm i} \partial_p
\Phi_{\rm i}$ at small $x$ are
$\partial_\mu \Phi^*_{\rm i} \partial_\nu \Phi^{\vphantom{*}}_{\rm i} 
= P_\mu P_\nu
\Phi^*_{\rm i} \Phi^{\vphantom{*}}_{\rm i} ,
$
because these contract with $q^\mu q^\nu$ to give a factor $s^2$.
Since $P^2$ is negligible, the dilaton and graviton-trace do not 
couple and so we
need the Laplacian for four-dimensional traceless symmetric tensors. 
Working out the
connection, one finds
\bel{lapla}
  \nabla^2 (P_\mu P_\nu  \Phi^*_{\rm i} \Phi^{\vphantom{*}}_{\rm i})
= P_\mu P_\nu e^{2A}  \nabla_0^2 (e^{-2A}  \Phi^*_{\rm i}
\Phi^{\vphantom{*}}_{\rm i}) \equiv P_\mu P_\nu  D^2
(\Phi^*_{\rm i} \Phi^{\vphantom{*}}_{\rm i})
\ ,
\ee
where $e^{2A}$ is a general warp factor as in \Eref{foursix}, $\nabla_0^2$ is
the scalar Laplacian, and we have defined
$ D^2 = e^{2A} \nabla_0^2 e^{-2A}$.

The exponential of $ \nabla^2$ is nonlocal, essentially a diffusion operator, with
$\ln x$ as diffusion time.  Consequently
the details of the geometry in the region where conformal symmetry is
broken can 
enter when the
diffusive ``time'' becomes large.  Still, we can
continue to use the
$AdS_5 \times W$ form~\eref{ffmurho} for the field strength, since its 
exponential falloff 
for $r < R^2 q$ means it is unaffected by the 
conformal symmetry breaking.
We then have
\bel{f12ex}
F_1  = \frac{\pi^2 \alpha' R^4 q^2}{ 4x^2} J_1
\ ,\quad
F_2  =\frac{\pi^2 \alpha' R^4 q^2}{ 2x} (J_0 + J_1)\ ,
\ee
where
\bel{Rconvo}
J_i = \int dw d^5\Omega \sqrt{\hat g_W}\, v^a v_a\, w^3 K_i^2(w)\, 
x^{-\alpha' D^2
/2} (\Phi^*_{\rm i} \Phi^{\vphantom{*}}_{\rm i})\ .
\ee
Omitting the correction term, one readily recovers the earlier 
result~\eref{Rw12}.

Now let us consider the asymptotic behavior as $x \to 0$.  The $x$ 
dependence will
be determined by the smallest eigenvalue $\lambda$ of $-\nabla^2$, of 
order $R^{-2}$, {\it i.e.}, $\alpha'\lambda \sim (gN)^{-1/2}$.  Thus we have
\bel{f12corr}
F_1 \propto x^{-2 + \alpha' \lambda/2}\ ,\quad F_2 \propto x^{-1 + \alpha'
\lambda/2}\ .
\ee
This is of the expected form~\eref{regexp}, and we identify the 
Pomeron intercept
as $2 - \alpha' \lambda/2$.  To get the $q$-dependence in this limit, 
we note that
the lowest normalizable eigenfunction of $\nabla_0^2$ falls as 
$r^{-4}$ as $r \to
\infty$, and so the lowest normalizable eigenfunction of $D^2$ falls 
as $r^{-2}$.
Then, up to $x$- and $q$-independent factors,
\bel{Rdimas}
J_i \propto x^{\alpha' \lambda /2} \int dw d^5\Omega \sqrt{\hat 
g_W}\, v^a v_a w^3
K_i^2(w)\, r^{-2}\ .
\ee
The integral is dominated by $w \sim 1$, where $r^{-2} \sim R^{-4} q^{-2}$.
Thus $F_{1,2}$ given by \Eref{f12ex} are independent of $q^2$ in this 
limit, and so
we will get indeed get a $q^2$-independent moment.

Before $x$ reaches the point where the conformal symmetry breaking region 
is felt by the diffusion operator, but at $x$ small enough that the $x^{-\alpha'\nabla^2/2}$
is not negligible, there is a rexgime where the diffusion occurs entirely in the 
AdS region.  In this case we can obtain a simple and interesting expression for the structure 
functions. In the AdS region,
\bel{d2}
D^2 = R^{-2}(\partial_u^2 - 4)\ ,\quad u = \ln (r/r_0)\ ,
\ee
and so
\bel{d2pp}
x^{-\alpha' D^2 /2}
(\Phi^*_{\rm i} \Phi^{\vphantom{*}}_{\rm i}) = x^{\alpha' \zeta / 2}\,
\Phi^*_{\rm i} \Phi^{\vphantom{*}}_{\rm i}\ ,\quad \zeta = \frac{4}{R^2} (1 -
\Delta^2) \leq 0\ .
\ee
Compared to the calculation in Section V.A, the only effect is a rescaling,
so \eref{Rw12} is slightly modified:
\begin{eqnarray}
F_1  &=&  x^{-(2 + \alpha' |\zeta|)/ 2}
\Biggl(\frac{\Lambda^2}{q^2}
\Biggr)^{\Delta - 1} \frac{\pi^2\rho |c_{\rm i}|^2}{  4 (4\pi gN)^{1/2}}
I_{1,2\Delta+3}\ , \ F_2=2x{2\Delta+3\over \Delta+2}F_1 \ .
\label{Rw12mod}
\end{eqnarray}
Note that since $\zeta < 0$, its effect on the structure functions
in the range where the 
form~\eref{Rw12mod} is valid makes them
grow {\it more} rapidly at small $x$.  The 
reason is simple.
The wavefunction $\Phi_{\rm i}$ is largest at small $r$, where the 
gauge field is
negligibly small.  The diffusion operator, which represents the 
logarithmic growth
of strings, allows this larger part of the wavefunction to interact 
with the gauge
field, and so increases the amplitude.  This effect shuts off, giving
the other behavior~\eref{f12corr}, when the diffusion reaches from the IR cutoff
$r_0$ all the way to the interaction radius $R^2 q$ --- this is at 
very small $x$,
corresponding to large diffusion time.  The
$\zeta<0$ growth is needed to connect the small-$x$ form, which falls 
with $q$, to
the
$q$-independent behavior at exponentially small $x$.

Finally let us evaluate the moments $M^{(i)}_2$.  Using the structure
functions~\eref{f12ex} yields
\begin{eqnarray}
M^{(1)}_1 &=& \frac{\pi^2 R^4 q^2}{ 2} \int dw d^5\Omega \sqrt{\hat 
g_W}\, v^a v_a\,
w^3 K_0^2(w)
\frac{1}{-D^2} \Phi^*_{\rm i} \Phi^{\vphantom{*}}_{\rm i}
\nonumber\\
M^{(2)}_2 &=& \frac{\pi^2 R^4 q^2}{ 2} \int dw d^5\Omega \sqrt{\hat 
g_W}\, v^a v_a\,
w^3 \Bigl(K_0^2(w) + K_1^2(w) \Bigr)
\frac{1}{-D^2} \Phi^*_{\rm i} \Phi^{\vphantom{*}}_{\rm i}
\ .
\label{momcalc}
\end{eqnarray}
This has a simple graphical interpretation.  The $1/D^2$ is a 
graviton propagator,
so this is the graph shown in Fig.~5.  The right vertex is the 
AdS/CFT representation
of the $J^\mu J^\nu T^{\rho\sigma}$ OPE coefficient, while the left vertex is
determined by the total energy-momentum of the hadron.  Thus we have 
reproduced  the
expected form.
\begin{figure}[t]
\begin{center}
\leavevmode
\epsfbox{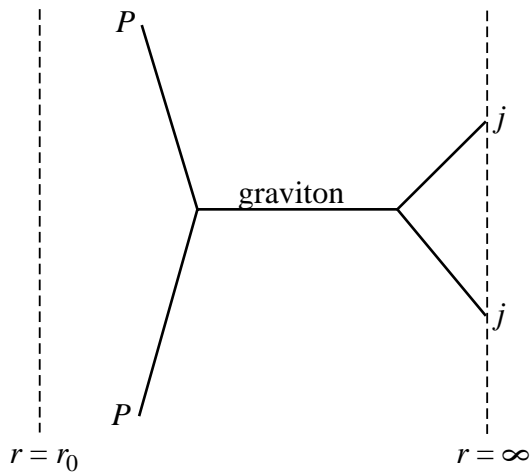}
\end{center}
\caption{Feynman graph giving the moment $M_2$.}
\end{figure}

To evaluate the moment explicitly, we must first solve
\bel{dsource}
-D^2 h = \Phi^*_{\rm i} \Phi^{\vphantom{*}}_{\rm i}
\ee
for $h(r,\Omega)$.  Integrating this equation over the interior 
of a sphere of radius 
$r_{\rm c}$
gives
\bel{dsint}
-\int_{r < r_{\rm c}} dr d^5\Omega \sqrt{-g}\, \nabla_0^2 (e^{-2A} h) =
\int_{r < r_{\rm c}} dr d^5\Omega \sqrt{g_\perp} e^{2A} \Phi^*_{\rm i}
\Phi^{\vphantom{*}}_{\rm i}\ .
\ee
Taking $r_{\rm c} \gg
r_0$, the right-hand side becomes the normalization integral~(\ref{normint})
and is equal to unity.  On the left-hand side, at $r \gg r_0$ the 
function $h$ goes
over to the normalizable AdS form for the graviton, proportional to 
$1/r^2$, with
the normalization fixed by~\Eref{dsint} to be
\bel{hsol}
h = \frac{1}{-D^2} \Phi^*_{\rm i} \Phi^{\vphantom{*}}_{\rm i} \to
\biggl( 4 r^2 R^2 \int_W d^5\Omega \sqrt{\hat g_W}  \biggr)^{-1}
\ .
\ee
As $q^2 \to \infty$, the Bessel functions cut off the integral~(\ref{Rconvo})
in the AdS region and so we can use the asymptotic form~(\ref{dsint}).  One
then finds 
\bel{momfin}
M_2^{(1)} \to \frac{\pi^2 }{5 }\frac{\displaystyle \int_W d^5\Omega \sqrt{\hat
g_W}
\,\hat
g_{W ab}\, v^a v^b} {\displaystyle \int_W d^5\Omega \sqrt{\hat g_W} }
\ ,\quad
\frac{M_2^{(2)}}{M_2^{(1)}} \to \frac{5}{3}\ .
\ee
Note that the moments are completely independent of the initial 
hadron.  They depend
only on the current, through the Killing vector $v^a$, which is 
consistent with their
relation to an OPE coefficient.  We conclude that our understanding 
of the distribution functions is complete, in the large $N$ limit.

Now let us understand from a somewhat more formal point of view the nature
of the string calculation that we are doing.  Because the size of the
excited strings is of order the curvature radius of the spacetime, we must
do a true string calculation, with vertex operators for the four external
states, on a sphere.  Three of these operators, say two gauge vertex
operators and one dilaton, can be fixed, and the world-sheet coordinate
$z$ of the remaining dilaton is integrated.  Because the spacetime
curvature is small in string units, the world-sheet coupling $(gN)^{-1/2}$
is also small.  However, there is another small parameter in the problem,
namely $x$, and we wish to keep corrections that are large when
$x^{(gN)^{-1/2}}$ differs significantly from unity.  In other words, we
want to keep all orders in $(gN)^{-1/2} \ln x$ while treating
$(gN)^{-1/2}$ itself as small.

This is reminiscent of the renormalization
group, and indeed it is precisely that on the world-sheet.  A standard
saddle point estimate shows that in the Regge regime the world-sheet path
integral is dominated small separation of the
two dilaton vertex operators,  $\delta z \sim \tilde t/\tilde s \sim x$.
We can then carry out the calculation in steps, the first of which is to
use the OPE to replace the product of dilaton vertex operators with a
single $t$-channel vertex operator (the operators that contribute are 
those of the
graviton Regge trajectory). To avoid large logs, this
operator itself must be renormalized at a scale comparable to the
separation
$\delta z$.  It is then necessary to run this operator to a reference
scale of order the radius of the sphere, and the final three-point
amplitude has no large logs.  The large logs of $x$ come only from the
running of the vertex operator.  This is precisely our correction factor
$x^{-\alpha' \nabla^2/2}$ --- the exponent is the world-sheet anomalous
dimension of the $t$-channel vertex operator.  This justifies our
prescription~\eref{ampmod}.

\section{Outlook}

Let us summarize the results of the string calculations, in particular with
regard to the different ranges of $x$.
\begin{description}
\item[A.]
At $x \gg (gN)^{-1/2}$ we have the supergravity results~(\ref{ffin}),
(\ref{fermfs}).
\item[B.]
At $x \ll (gN)^{-1/2}$
but with $|\ln x| \ll (gN)^{1/2}$ we have the form~(\ref{Rw12}),
where the excited strings form a continuum.
\item[C.]
When $|\ln x| (gN)^{-1/2}$ becomes
of order one we must take into account the effect of string growth, 
leading to the
form~\eref{Rw12mod}.
\item[D.]
Finally, when $|\ln x| (gN)^{-1/2} > \ln ( \Lambda/q)$ we go over to the final
asymptotic form~\eref{f12corr}.
\end{description}
The behavior B is actually a special case of C, in that the
form~\eref{Rw12mod} reduces to (\ref{Rw12}) when $|\ln x| 
(gN)^{-1/2}$ is small.
The transition between forms A and B is described by a discrete sum 
over excited
strings,~\Eref{w12}.  The transition between C and D would similarly 
be described
by a discrete sum over eigenvalues of $D^2$.

In region A, we have found a transition at $gN\sim1$
--- more precisely, a crossover ---
from partonic to hadronic behavior.  This crossover
is interesting from a number
of points of view.  For one thing, it is related to a
string theory crossover, in which a set of $N$ D-branes
with small string coupling should be replaced,
when $gN\gg 1$, by a black brane with a large gravitational field
and a horizon.  There are important long-standing questions about
this mysterious
transition.  Where did the branes, and the open
strings on their world-volume, go?  Which aspects of the
physics are the same on both sides of $gN\sim1$, and which
are fundamentally different?  Along the same lines is the question
of precisely how, for any computation in this background, a result calculated
on worldsheets with boundaries
and holes at small $gN$ can be calculated
with worldsheets with no holes at large $gN$; how do the holes
close, or become subleading?

Although we have not concretely answered these questions, we have
related them to a system which is easier to understand --- the
operator spectrum of the world-volume field theory. We have done so using nonperturbative
methods in field theory, making it possible to qualitatively follow
the transition from small to large $gN$.  The shifting of operator
dimensions relative to one another is easy to understand and describe,
even at the point of transition where the physics is complicated in
both perturbative field theory and in perturbative string theory.

Using this knowledge one can roughly understand which classes of
Feynman graphs, or more precisely, which regions of kinematics within
those Feynman graphs, are suppressed as $gN\to1$.  In this regime, the
probability for parton emission is so great, for any colored line
whose color is not immediately cloaked, that any hole in the
worldsheet --- any place where a color index is carried by something
physically separated from the corresponding anticolor --- becomes
densely filled with multiple partons.  Note this has nothing to do
with confinement; there are no flux tubes here.  It is simply that the
probability that a light color charge will radiate other light colored
particles is so great that no such particle will ever be isolated.
Thus it seems the holes in the worldsheet --- the holes between
colored particles in Feynman graphs --- fill in not because they
shrink (with the gluons, thin ribbons in 't Hooft's double-line
formalism, somehow becoming thick) but because they are broken up into
tiny fragments, as
though covered over by cheesecloth, 
as a myriad of additional thin ribbons are emitted.  To localize and scatter off a
parton with substantial $x$ inside a hadron requires separating it
from its neighbors; this will not happen in the large $gN$ regime.
Again, we stress this has nothing to do with confinement; this is also
clear from the fact that \nfour\ Yang-Mills is described by closed
strings at large $gN$.

 From a particle physicist's point of view, the interest of these
results lies in two places.  On the one hand, they gives some
conceptual insights into QCD.  We have
shown that, at large $gN$, DIS in the moderate $x$ region
is quite different from QCD.
This is true despite the fact that various scaling laws, such as
the ``parton-counting" rules \cite{bf,mmt}, are still valid
in this regime \cite{PShard}, since they follow from (approximate)
conformal invariance.  This suggests that it may be useful to
investigate still further the conformally invariant aspects of QCD
\cite{QCDcft} in order to cleanly separate its dynamics from its kinematics.
Still, DIS in QCD at moderate $x$ is well understood and our
results, though interesting, are not likely to suggest significant technical
improvements.

On the other hand, although QCD is not at large 't Hooft parameter, there may exist
gauge interactions in nature which
are, perhaps associated with electroweak
symmetry breaking.  Indeed, the dual string description of such a theory is
precisely Randall-Sundrum
compactification~\cite{rsone}. Various
particles of the standard model may be composite at 1 TeV,
with others being pointlike at that scale.   In this case,
DIS may actually occur at LHC, or more likely, at VLHC.
The scattering of a pointlike particle off
of a composite one will be described as DIS of the sort
we have just described, and the absence of hard partons inside
the composite object will have calculable and measurable
experimental consequences.

At small $x$, Regge physics, in the form of $t$-channel
Pomeron exchange summarizing
the effect of large numbers of hadronic resonances in the $s$-channel,
appears both at small $gN$ (as evidenced
both by calculation in QCD and by HERA data) and at large $gN$.  One
difference is that the Pomeron intercept is
close to 1 at small $gN$ and close to 2 at
large $gN$.  (We will not address the controversy regarding hard versus
soft Pomerons in QCD.)  However, there is still a transition at $gN\sim 1$.
The Pomeron scatters off a parton at small $gN$, and has very little
$q^2$ dependence, while it scatters off the entire hadron at large $gN$ (or
off a hadron in the cloud surrounding the parent hadron) and its
contribution falls as 
a power of $q^2$.

As $x$ becomes extremely small, physical
QCD is affected by multiple Pomeron exchange, as is the string theory;
addressing this physics requires methods beyond those discussed
here.  However, for $N$ exponentially large the situation simplifies somewhat.
Here we find the most notable aspect of the string calculations: the
use of the world-sheet renormalization group to include the effect of the
growth of strings.  (A related use of the renormalization group appeared in
Ref.~\cite{size}.)  It is this nonlocal behavior, which involves a form of diffusion from 
small distances to large which is not completely unfamiliar in QCD problems
with very different length scales, which permits the $q^2$-independent
energy-momentum sum-rule to be satisfied in region D.  For region B,
with its substantial $q^2$ dependence, to match on to region D requires
the intermediate C region, where the nonlocality first begins to play a role.

The high-energy growth of strings is one of their most
distinctive non-field-theoretic properties, and it has been argued for
example to be associated with the resolution of the black hole information
problem~\cite{growth}.  It is an interesting direction to investigate 
whether the
correction that we have discussed, in addition to possible conceptual
applications in QCD, has any relevance in black hole or
cosmological contexts.


\vskip 1.0in

We would like to thank Stan Brodsky, 
Leonard Gamberg, Al Mueller, Michael Peskin, and Lenny
Susskind for many valuable comments and discussions.  We would also
like to acknowledge interesting discussions with Oleg Andreev, John Collins and Hans Pirner.
The work of J.P. was supported by National Science
Foundation grants PHY99-07949 and PHY00-98395; that of M.J.S.  was
supported by DOE grant DOE-FG02-95ER40893, NSF grant PHY-0070928, and
by the Alfred P. Sloan Foundation.  M.J.S. thanks the Institute for
Advanced Study and the Aspen Center for Physics where much of this
work was done.



\end{document}